\documentclass[11pt,tightenlines,eqsecnum,floats,aps,amssymb,nofootinbib,prd,shownopacs,floatfix,superscriptaddress]{revtex4-2}

\usepackage[T1]{fontenc}
\usepackage[utf8]{inputenc}

\usepackage{amsmath,amssymb,amsthm, mathtools,bbm, dsfont, mathrsfs}
\usepackage{tensor, braket}
\usepackage{mlmodern}
\usepackage{xparse}
\usepackage{color}
\usepackage{xfrac}
\usepackage[dvipsnames]{xcolor}

\usepackage{parskip}
\usepackage{microtype}
\usepackage{booktabs}
\usepackage{empheq}
\usepackage[a4paper,  left=30mm,  right=30mm, top=30mm, bottom=30mm, footnotesep=0.3cm]{geometry}
\usepackage{fancyhdr}
\usepackage[most]{tcolorbox}
\usepackage{longtable,booktabs}
\usepackage[colorinlistoftodos,prependcaption,textsize=tiny]{todonotes}
\usepackage{float}

\usepackage{graphicx}
\usepackage{epstopdf}
\usepackage{latexsym}
\usepackage[normalem]{ulem}
\usepackage{dsfont}
\usepackage[toc,title,page]{appendix}
\usepackage[sort&compress]{natbib}
\usepackage{etoolbox}

\usepackage{tikz}
\usepackage{enumitem}

\usepackage{hyperref}
\makeatletter
\newcommand{\sumint}{\mathop{\mathpalette\sumint@\relax}\slimits@}
\newcommand{\sumint@}[2]{\ooalign{$\m@th#1\sum$\cr\hidewidth$\m@th#1\int$\hidewidth\cr}}
\makeatother

\newcommand{\C}{\mathcal{C}}
\renewcommand{\S}{\mathcal{S}}
\newcommand{\E}{\mathcal{E}}
\newcommand{\SE}{\mathcal{SE}}
\newcommand{\CE}{\mathcal{CE}}

\newcommand{\Id}{{\mathds{1}}}

\newcommand{\CH}{C_\mathrm{tot}}
\newcommand{\CO}{C_\mathrm{0}}
\newcommand{\Hint}{H_{\mathrm{int}}}

\newcommand{\rkin}{\rho^{\mathrm{kin}}}
\newcommand{\rphys}{\rho^{\mathrm{phys}}}

\theoremstyle{remark}\theoremstyle{plain}

\makeatletter
\renewcommand{\p@subsection}{}
\renewcommand{\p@subsubsection}{}

\begin{document}
\begin{abstract}
We derive a relational master equation for the system's density matrix from a restricted class of microscopic open quantum mechanical models in which the quantum clock interacts with both the system and environment. To achieve this in the perspective-neutral approach, we generalize the construction of the clock POVM by means of a spectral dressing operator that relates the free and interacting theories, allowing us to construct a fully covariant POVM that is adapted to the interactions. This is then used to define a generalized Page-Wootters conditioning and obtain a relational state whose dynamics and clock reading are influenced by the interactions. Furthermore, we show that if the POVM is not adapted to the interactions, in general, this can yield situations in which the relational dynamics and quantities derived from it become gauge-dependent, which is not the case if we use a POVM adapted to the interactions. We find four distinct contributions in the relational master equation due to the quantum clock: a modified effective Hamiltonian that involves the one-point clock correlation function, joint clock-environment correlation functions, yielding decoherence and dissipation, spectral dressing correction encoding backreaction of the quantum clock, and an inhomogeneous term due to possible initial system-environment correlations. We investigate these effects in three examples, including a non-interacting clock as a specific limiting case, an ideal interacting clock and a qubit clock coupled to a qubit system. At the microscopic level, whether the reduced dynamics are Markovian depends crucially on the details of the interactions and the clock states used to construct the POVM.
\end{abstract}
\title{Relational open quantum systems: Master equations with interacting quantum clocks from microscopic models}
\author{Kristina Giesel}
\email{kristina.giesel@fau.de}
\affiliation{Institute for Quantum Gravity, Theoretical Physics III, Department of Physics,  Friedrich-Alexander-Universit\"at Erlangen-N\"urnberg, Staudtstr. 7, 91058 Erlangen, Germany.}

\author{Ashay Sathe}
\email{ashay.sathe@fau.de}
\affiliation{Institute for Quantum Gravity, Theoretical Physics III, Department of Physics,  Friedrich-Alexander-Universit\"at Erlangen-N\"urnberg, Staudtstr. 7, 91058 Erlangen, Germany.}
\maketitle
\newpage

\section{Introduction}
\label{sec:Intro}
Unlike closed quantum systems, open quantum systems take into account that no physical system is ever perfectly isolated but interacts, at least weakly, with its environment \cite{Breuer:2002,Lindblad:1975ef,Gorini:1975nb,nakajima_quantum_1958,zwanzig_ensemble_1960}.
The environment contains all relevant degrees of freedom of the total system except those of the system of interest. The dynamics of the system's density matrix under the effective influence of the environment can be described by a master equation. One obtains this master equation by tracing out the environmental degrees of freedom in the von Neumann equation of the total system including system and environment. Such a master equation contains, in addition to the unitary dynamics, which is also present in closed systems, a dissipator that encodes decoherence and dissipation due to the coupling to the environment.
In most existing works in the literature, the evolution of the master equation is formulated with respect to an external and classical time parameter. However, in real experiments, time is usually not read off from some external parameter but from a physical system, e.g., an atomic clock, a harmonic oscillator, a pointer on a dial, etc., with its own dynamics being a part of the total physical system. In this case, we split the total system into a clock and the rest, consisting of the system of interest and the environment, and formulate the dynamics of the non-clock degrees of freedom with respect to the chosen physical clock. This raises the question of how the master equation and the resulting decoherence and dissipation are modified when time is treated relationally, as read off from such physical clocks.

This perspective, in which the clock is part of the system, is adopted by the relational formalism \cite{Rovelli:1990pi,Rovelli:1990ph,Vytheeswaran:1997jr,Dittrich:2004cb,Dittrich:2005kc,Thiemann:2004wk}. At the classical level this corresponds to promoting the clock to a kinematical degree of freedom on the same footing as the system and environment. 
Any classical model with an external time parameter can be reformulated in a reparameterization-invariant way through this promotion. To do so, one extends the kinematical phase space by the clock degree of freedom and its conjugate momentum. The dynamics are then formulated relative to the chosen clock rather than to the external time parameter.
Requiring that the physics does not depend on an arbitrary relabeling of the clock's reading implies that the total Hamiltonian vanishes, i.e., one has a Hamiltonian constraint in the model. This is the structure that arises naturally in theories without a background time, such as general relativity. It can, however, be imposed on any system to make the relational description explicit. In such a setting all dynamics is relational. This means that the meaningful temporal statements in this context are of the form ``what is the state of the system and the environment when the clock reads a specific value?''. The physical degrees of freedom are encoded in Dirac observables, which are quantities that Poisson commute with the Hamiltonian constraint. There exists a systematic way to construct Dirac observables, even in the more complicated setup of general relativity, once a clock (or, more generally, a physical dynamical reference frame) is chosen \cite{Vytheeswaran:1997jr, Dittrich:2004cb,Dittrich:2005kc,Thiemann:2004wk}. When the clock is treated classically, one obtains the physical sector of the quantum theory by applying a reduced phase space quantization in which only the physical phase space from the perspective of the classical clock is quantized. In this approach, the constraint is completely solved at the classical level. After quantization, one then has a true physical and non-vanishing Hamiltonian operator that generates the evolution of the quantized Dirac observables. Following this, the standard derivation of the master equation then carries over to the relational setting. This yields a master equation for the system's density matrix as a function of the relational clock time rather than the external time parameter (see, for instance, \cite{Fahn:2022zql,Fahn:2026pov} for an application in linearized gravity where geometrical clocks have been chosen).

The main question we want to address in this work is how the master equation and the resulting decoherence and dissipation are modified when time, instead of being an external parameter, is treated relationally with quantum clocks which, in general, interact with the system and environment. This will also allow us to recover the case of the classical clock in a certain limit. In general, we expect that quantum clocks yield additional decoherence and dissipation when compared to classical clocks. This has been shown in \cite{Gambini:2006yj} for instance, where a phenomenological model was used to describe the quantum clocks and also in more recent work in \cite{Luppi:2026qyv} where a non-interacting clock is treated as a quantum reference frame. A recent complementary approach in \cite{Nandi:2026muw} addresses related questions in a relativistic setting, as we discuss in more detail below.

The aim of this work is to provide a framework in which these effects can be investigated in detail for quantum mechanical models. The methods and techniques introduced in this article cannot yet cover the most general model. Instead this should be understood as a first step towards systematically deriving and investigating relational open quantum models with interacting quantum clocks from microscopic models as a basis for future work wherein we want to generalize the methods and techniques in various directions. 

The implementation of quantum clocks in the relational formalism for closed systems is well established for quantum mechanical models in the context of the perspective-neutral approach \cite{Vanrietvelde:2018pgb,Hohn:2018toe,Hoehn:2019fsy,delaHamette:2021oex,AliAhmad:2021adn,Giacomini:2017zju} (see also~\cite{Aharonov:1967zza,Aharonov:1984zz,Bartlett:2006tzx} for earlier and alternative formulations of quantum reference frames). More recent work in the context of quantum field theories for linearized gravity using ideal clocks can be found in \cite{Chen:2026kui} and for gravitational null rays with non-interacting clocks in \cite{Freidel:2026gcp}. Instead of reduced phase space quantization, one applies Dirac quantization \cite{Dirac1964,Henneaux:1992ig,Thiemann:2007pyv} and quantizes the entire kinematical phase space, yielding the kinematical Hilbert space that includes the clock degrees of freedom. The classical constraint is promoted to an operator on the kinematical Hilbert space. In the case of non-interacting clocks, which have mainly been considered so far in the literature, the clock reading can be described following the spirit of Page-Wootters \cite{Page:1983uc,Wootters:1984wfv,Giovannetti:2015qha,Marletto:2016gwv,Hoehn:2019fsy}. This involves constructing a positive operator valued measure (POVM) associated with the clock that can be used to obtain a clock reading in terms of a conditional trace. The clock POVM is required to transform covariantly under the finite transformation the constraint operator generates via its unitaries. In the case of non-interacting clocks this simplifies to the requirement of covariance under the finite transformation generated only by the free clock Hamiltonian \cite{Page:1983uc,Hoehn:2020epv,Hoehn:2019fsy}. 
Kinematical density matrices can be mapped to physical ones by applying the $\mathcal G$-twirl \cite{Bartlett:2006tzx,Loveridge:2016tnh,Krumm:2020fws,Castro-Ruiz:2021vnq,Kitaev:2003zj,Spekkens:2008acy,Bartlett:2003ryf,Loveridge:2017pcv,Carette:2023wpz}, which is a special case of group averaging \cite{Ashtekar:1995zh,Giulini:1998rk,Marolf:2000iq}. This allows us to extract the physical sector in the quantum theory. By construction, physical quantities then commute with the Hamiltonian constraint operator, mirroring the classical property that Dirac observables Poisson commute with the Hamiltonian constraint.

While the perspective-neutral framework for quantum clocks is well developed for non-interacting clocks, the assumption of non-interacting clocks becomes particularly restrictive in the context of open quantum systems. 
To begin with, there is no obvious reason why the clock should not couple to the environment like the system does. This becomes particularly clear if one regards choosing a clock as choosing a subsystem within a larger system. For some such clock choices, the clock would inevitably couple to the environment.
Moreover, in general beyond open system settings, clocks will also couple to the part we identify as the system. This is, in particular, the case if we consider models involving gravity, since everything couples at least gravitationally.
For instance, the matter reference fields constructed from scalar or dust fields, which are used in the context of general relativity and its quantization, couple gravitationally even in the case of closed systems \cite{Kuchar:1990vy,Brown:1994py,Giesel:2007wn,Giesel:2012rb,Domagala:2010bm,Giesel:2016gxq,Husain:2011tk}. The existing literature on interacting quantum clocks has so far focused on closed systems \cite{Zych:2011hu,Castro-Ruiz:2017whw,Smith:2019imm,Paige:2018lrd,Castro-Ruiz:2019nnl}. While these works established important results on quantum clock-induced modifications of the Schr\"odinger equation, they do not address the open system setting nor do they derive master equations from microscopic models. This motivates extending the relational master equation framework to the case of interacting quantum clocks, which is the central goal of this work.

In general, interactions of the chosen quantum clock become relevant in two ways. First, they change the dynamics of the density matrix of the combined system-environment. Second, they also affect the quantum measurement through which we define the clock reading. To account for the second effect, we introduce a spectral dressing operator which relates the interacting theory to the free theory and that can be constructed in the restricted class of models that we consider in this work. The class of models we consider is characterized by the existence of a well-defined spectral dressing operator, which requires that the eigenstate structure of the constraint operator varies continuously with the coupling, i.e., no level crossings, no bound states appearing or disappearing, and the invertibility of a certain superoperator entering the kernel of the master equation throughout the evolution. These are the analogs, for the clock sector, of the weak coupling and invertibility assumptions underlying the standard time-convolutionless (TCL) derivation of the usual master equation if the environmental degrees of freedom are traced out, and here they define the regime in which a systematic perturbative treatment of interacting quantum clocks is currently available. This spectral dressing operator also enters into the construction of a generalized clock POVM which is covariant under the full Hamiltonian constraint, rather than only under the free clock Hamiltonian as in existing treatments. This generalized POVM reduces to the standard non-interacting POVM when interactions are switched off, and allows us to formulate a generalized Page-Wootters conditioning that is adapted to the interacting setting. The resulting relational master equation that describes the evolution of the system's density matrix contains three distinct contributions absent in the standard case: a modification of the effective Hamiltonian through the mean value of the clock coupling, a dissipator sourced by genuine quantum fluctuations of the clock that modifies decoherence and dissipation, and a correction arising from the spectral dressing that encodes the backreaction of the interacting clock on the relational time reading. In addition, an inhomogeneous contribution arises from possible initial system-environment correlations generated by the ${\cal G}$-twirl. We recover the standard relational master equation in the limits of non-interacting and ideal clocks, and illustrate the general results through three examples of increasing complexity.

As we will also show in this work, constructing a clock POVM that is adapted to the interactions in the model is not just a technicality. In specific cases, it can be consistent to use a clock POVM that is covariant only under the free clock evolution even when interactions are present, as done in \cite{Smith:2017pwx} for instance. However, in general this can yield situations in which the relational dynamics and physical quantities derived from it become gauge-dependent. In one of the examples involving a qubit clock coupled to a qubit system that is discussed later, this leads to coherences that are constant in one gauge and oscillate in another gauge. In contrast, the form of coherences computed based on a relational state with respect to a fully covariant POVM does not depend on the chosen gauge. 

A recent and important contribution to relational master equations can be found in \cite{Nandi:2026muw}, where they develop a covariant relational framework for relativistic open quantum field theories and derive a covariant master equation on spacetime hypersurfaces using the Schwinger-Tomonaga equation. Herein, relational time is defined through a finite resolution clock instrument. The main result presented in their work is that finite clock resolution generically induces non-Markovian reduced dynamics. They show that the familiar structure of the master equation emerges only after a relational coarse-graining. 
The present work is complementary to this approach. 
While \cite{Nandi:2026muw} establish the relativistic covariance of relational open quantum field models and the connection to the gravitational decoherence model in \cite{Anastopoulos:2013zya}, they characterize the clock through a finite resolution instrument operationally. In this work we consider quantum mechanical models in the perspective-neutral approach and provide a fully microscopic derivation. In particular, for interacting quantum clocks we construct the POVM explicitly using a spectral dressing operator, include a microscopic clock-system-environment interaction, and derive the joint clock-environment correlation functions without assuming statistical independence of the clock and environmental fluctuations as done in \cite{Nandi:2026muw}. This allows us to identify and characterize the three distinct quantum clock-induced contributions to the relational master equation mentioned above.

The paper is structured as follows: After the introduction in section~\ref{sec:Intro}, section~\ref{sec:Review} contains a brief review of relational open quantum systems with non-interacting clocks wherein we present the methods and techniques relevant for the further sections in this article. The usual treatment of open quantum systems with an external time parameter is discussed in subsection~\ref{sec:OQS}, while subsection~\ref{sec:rel-tcl-crfs} considers relational open quantum systems with classical clocks. In subsection~\ref{sec:qclocks-rev} the existing results on non-interacting clocks in the context of closed quantum systems which are relevant for this work are reviewed. Section~\ref{sec:relational-master-equation} focuses on the central aim of this work, which is to derive a relational master equation with interacting quantum clocks. This is done in several steps: First, in subsection~\ref{sec:GenPOVMInteracting} a generalized clock POVM is constructed by means of a spectral dressing that is adjusted to the interactions of the clock. Second, in subsection~\ref{sec:RelStateIntClocks} this POVM along with the ${\cal G}$-twirl is used to define a generalized Page-Wootters conditioning, which yields the relational state for the system-environment. The corresponding dynamics of this relational state and thus the final relational master equation with interacting quantum clocks are then derived in subsection~\ref{sec:DerRelMasterEqn}. Section~\ref{sec:examples} discusses applications of the relational master equation which is derived in section~\ref{sec:relational-master-equation} via three specific examples: the non-interacting clock in subsection~\ref{sec:ex-nonint}, the interacting ideal clock in subsection~\ref{sec:ex-ideal}, and a qubit clock coupled to a qubit system as an example of a non-ideal interacting quantum clock in subsection~\ref{sec:ex-qubit}. Finally, in section~\ref{sec:Conclusion} we summarize our results, draw conclusions and outline directions for future research.
\section{Brief review on (relational) open quantum systems}
\label{sec:Review}
In this section we briefly review relational open quantum systems with a focus on the methods and techniques relevant for the work in the remaining sections of this article.
\subsection{Open quantum systems and the time-convolutionless master equation}\label{sec:OQS} 
In this subsection, we briefly summarize some standard results from the theory of open quantum systems which we use in the remainder of this work. More detailed derivations can, for instance, be found in the textbook \cite{Breuer:2002}. For a more detailed presentation adapted to the notation used here, we refer the reader, for example, to \cite{Fahn:2022zql}.\footnote{For brevity, we omit hats on operators.}

\begin{figure}[!h]
  \centering
    \includegraphics[width=0.7\linewidth]{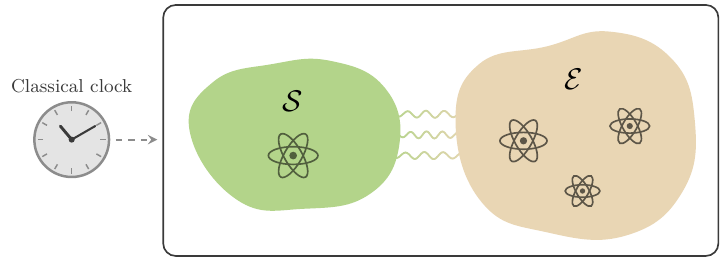}
\caption{An illustration of a conventional open quantum system: system $\S$ and environment $\E$ interact, with time given by an external classical clock.}
\label{fig:oqs}
\end{figure}

We consider the approach to open quantum systems where one starts with a total closed system that can be decomposed into a system of interest $\S$ coupled to an environment $\E$ as illustrated in Fig.~\ref{fig:oqs}. Their total dynamics is described via a Hamiltonian operator on the Hilbert space ${\cal H}_{\rm tot}={\cal H}_{\S}\otimes {\cal H}_{\E}$ of the form
\begin{equation*}
  H_{\rm tot} = H_{\S}\otimes\Id_{\E} + \Id_{\S}\otimes H_{\E}+\Hint
  =H_{\S}\otimes\Id_{\E} + \Id_{\S}\otimes H_{\E}
    + \sum_{j}\alpha_j\,O_{\S,j}\otimes O_{\E,j}\,,
\end{equation*}
where $H_{\S}$ and $H_{\E}$ are the free Hamiltonians of $\S$ and $\E$ respectively, and $\Hint$ denotes the interaction Hamiltonian. In this work we assume that $\Hint$ consists of self-adjoint system and environmental operators $O_{\S,j}$ and $O_{\E,j}$ respectively. The constants $\alpha_j\in\mathbb{R}$ denote the coupling constants for the individual contributions of the interaction.

In order to derive a master equation for the system, we choose an initially uncorrelated state $\rho(0)=\rho_{\S}(0)\otimes\rho^{\mathrm{ref}}_{\E}$ with a stationary reference state $\rho^{\mathrm{ref}}_{\E}$ of the environment such that $[H_{\E},\rho^\mathrm{ref}_\E]=0$. Performing the environmental partial trace of the von Neumann equation for $\rho_{\rm tot}$, one obtains the time-convolutionless master equation at second order in the couplings (TCL$_2$) for the reduced system state $\rho_\S(t)$ that reads
\begin{align}
  \frac{\mathrm{d}}{\mathrm{d}t}\rho_{\S}(t)
={}&-\frac{i}{\hbar}
\big[H_{\mathcal S},\rho_{\mathcal S}(t)\big]
\nonumber\\
&-\frac{1}{\hbar^2}
\sum_{j,k}\alpha_j\alpha_k
\int_0^t \mathrm{d}s\,\Big(
C_{jk}(t-s)
\big[O_{\mathcal S,j},
\widetilde O_{\mathcal S,k}(s-t)
\rho_{\mathcal S}(t)\big]
\nonumber\\
&\hspace{32mm}
+C_{jk}^*(t-s)
\big[\rho_{\mathcal S}(t)
\widetilde O_{\mathcal S,k}(s-t),
O_{\mathcal S,j}\big]
\Big)
+\mathcal O(\alpha^3),
\label{eq:TCL2}
\end{align}
where $\widetilde{O}_{\S,k}( s-t )=e^{\frac{i}{\hbar}H_{\S} (s-t) } O_{\S,k}\,e^{-\frac{i}{\hbar}H_{\S} (s-t) }$, analogously for $\widetilde{O}_{\E,j}( s-t )$ with $H_{\E}$, and the environmental correlation functions are given by
\begin{equation*}
  C_{jk}( t-s )=\operatorname{Tr}_{\E}\big(\widetilde{O}_{\E,j}( t -s )\,
    O_{\E,k}\,\rho^{\mathrm{ref}}_{\E}\big)\,.
\end{equation*}
Note here that we assumed that the model has vanishing environmental one-point functions $\operatorname{Tr}_{\E}(O_{\E,j}\rho^{\mathrm{ref}}_{\E})=0$. If this is not given a priori, we can redefine $O_{\E}$ appropriately to ensure this.

The first term in \eqref{eq:TCL2} encodes the unitary evolution. The second term is the dissipator which can encode a Lamb shift contribution that modifies the unitary evolution, as well as decoherence, and dissipation. 

To arrive at the \textit{Lindblad form} of the master equation \cite{Lindblad:1975ef,Gorini:1975nb}, two approximations are made in addition to the assumptions above, namely a Markov approximation and a rotating wave approximation. In the Markov approximation, the correlation functions $C_{jk}( s )$ are assumed to decay on a time scale shorter than the relaxation time of $\S$. This allows us to send the upper limit of the temporal integral in \eqref{eq:TCL2} to infinity. As a consequence the coefficients become time-independent. The second approximation is the rotating wave approximation (RWA) \cite{Davies:1974dcc,Breuer:2002,Lindblad:1975ef,Gorini:1975nb} in which rapidly oscillating terms with different Bohr frequencies are neglected. The master equation then takes the Lindblad form
\begin{equation}
\frac{\mathrm{d}}{\mathrm{d}t}\rho_{\mathcal S}
=-\frac{i}{\hbar}[H_{\mathcal S}+H_{\rm LS},\rho_{\mathcal S}]
+\sum_k\gamma_k
\left(
L_k\rho_{\mathcal S}L_k^\dagger
-\frac12\{L_k^\dagger L_k,\rho_{\mathcal S}\}
\right),
\end{equation}
where $H_{\rm LS}$ is the above-mentioned Lamb shift contribution that shifts the unitary energies. Here, the non-negative coefficients $\gamma_k$, also referred to as the decay rates, are determined by the environmental correlation functions. The Lindblad operators $L_k$ are constructed from linear combinations of the Bohr-frequency components of the system coupling operators $O_{\S,j}$ with coefficients determined via diagonalization of the corresponding decay rate matrices.

\subsection{Relational TCL master equation with classical clocks}\label{sec:rel-tcl-crfs}
\begin{figure}[H]
  \centering
    \includegraphics[width=0.5\linewidth]{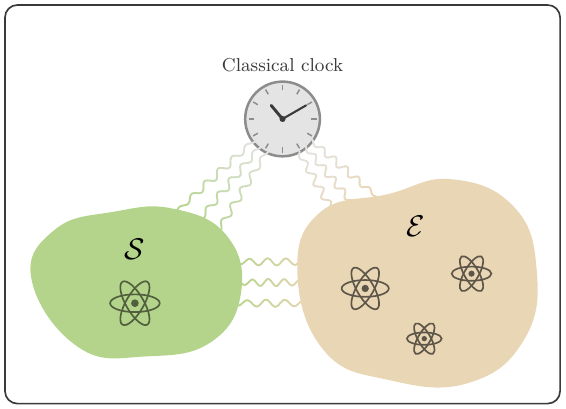}
\caption{An illustration of a relational description with a classical clock: the clock is part of the total system and interacts with $\S$ or $\E$ but stays classical in the formulation of the quantum model.}
\label{fig:RelClassicalClock}
\end{figure}

In the previous subsection wherein we discussed the usual formulation of open quantum systems, time entered as an external classical time parameter. Now, we briefly review the relational formulation of open quantum systems with classical reference frames illustrated in Fig.~\ref{fig:RelClassicalClock}. This will allow us, later, to compare the limit of ideal quantum clocks with the case of classical clocks in detail. 
We can reformulate any given classical system with an external time parameter into a relational formulation \cite{Rovelli:1990pi}. To do this, we can extend the kinematical phase space by an additional canonical pair $(T,P_T)$ associated with the classical clock and its momentum. We further assume that, in general, the clock interacts with $\SE$, which means that $\Hint$ can also have a dependence on $T$. This allows us to formulate the given system in a reparameterization-invariant way. The usual Hamiltonian of the environment and the system is enlarged to a first class constraint that now involves the clock momentum linearly and contains additional contributions of the clock in the interaction terms,
\begin{equation*}
 \CH = P_T + H_{\mathrm{tot}} \approx 0,\quad H_{\mathrm{tot}} = H_{\S}(q^i_{\S},p_i^{\S}) + H_{\E}(q^i_{\E},p_i^{\E}) + \alpha \Hint(T,q^i_{\S},p_i^{\S},q^i_{\E},p_i^{\E}).   
\end{equation*}
Even at the classical level, the reference frame (here the clock) interacts with the other degrees of freedom in general.
The physical phase space that includes the physical degrees of freedom consists of the Dirac observables of all degrees of freedom apart from those of the clock. These Dirac observables are quantities that Poisson commute with the constraint $\CH$. They can be constructed using the methods developed in \cite{Vytheeswaran:1997jr,Dittrich:2004cb,Dittrich:2005kc,Thiemann:2004wk}. They can be expressed as a power series in the clock, with coefficients constructed from nested Poisson brackets of the constraint and the phase space function whose observable we aim to construct. 
We denote the Dirac observables of the kinematical elementary variables by ${\cal O}_{T,\S}:=({\cal O}_{q^i_{\S},T},{\cal O}_{p_i^{\S},T})$ and ${\cal O}_{T,\E}:=({\cal O}_{q^i_{\E},T},{\cal O}_{p_i^{\E},T})$, where the label $T$ emphasizes that they are constructed with respect to the clock $T$. Their dynamics are generated by a physical Hamiltonian that, at the classical level, has the form $H_{\rm phys}=H_{\rm tot}({\cal O}_{T,\S},{\cal O}_{T,\E},T=\tau)$ where $\tau$ denotes the clock reading. Here we have used that ${\cal O}_{T,T}=\tau$. The Dirac observable ${\cal O}_{q^i_{\S},T}$ gives the value of $q^i_\S$ when the clock has the value $\tau$, and analogously for all other constructed observables. Having done this, we have now reformulated the originally constrained system as an unconstrained relational Hamiltonian system for the physical degrees of freedom. 

To obtain the quantum model and the relational master equation we now apply a reduced phase space quantization. The underlying algebra that is considered as a starting point for canonical quantization is the Poisson algebra of the sets of Dirac observables ${\cal O}_{T,\S},{\cal O}_{T,\E}$. In the case considered here, they satisfy standard canonical brackets and are implemented as multiplication and derivative operators on the perspectival physical Hilbert space ${\cal H}^T_{\rm phys}={\cal H}_{\S}\otimes {\cal H}_{\E}$, where we introduced the label $T$ to emphasize that it is from the perspective of the clock $T$. The relational time-dependent Hamiltonian $H^{\rm phys}$, that encodes the classical dynamics, is promoted to a time-dependent Hamiltonian operator on ${\cal H}^T_{\rm phys}$, given by
\begin{equation*}
  H^{\rm phys} = H^{\rm phys}_{\S}\otimes\Id_{\E} + \Id_{\S}\otimes H^{\rm phys}_{\E}+\Hint^\mathrm{phys},
\end{equation*}
with
\begin{equation*}
\Hint^\mathrm{phys}=  \sum_{j}\alpha_j f(\tau)\,O^{\rm phys}_{\S,j}\otimes O^{\rm phys}_{\E,j}\,,
\end{equation*} 
where the function $f$ is determined by the explicit dependence of $H^{\rm phys}$ on $\tau$ at the classical level. The setup is now similar to the one in subsection \ref{sec:OQS} with the main difference being that the quantum dynamics generated by $H^{\rm phys}$ is not with respect to some external time $t$ but rather with respect to the classical relational time $\tau$. We can now proceed, in a similar way as before, to obtain the relational TCL$_2$ equation by tracing out the physical environmental degrees of freedom. As before, assuming an initially uncorrelated state $\rho_\SE=\rho_\S\otimes\rho_\E^{\rm ref}$ we get
\begin{align}
  \frac{\mathrm{d}}{\mathrm{d}\tau}\sigma^{\rm phys}_{\S}(\tau)
={}&-\frac{i}{\hbar}
\big[H^{\rm phys}_{\mathcal S},\sigma^{\rm phys}_{\mathcal S}(\tau)\big]
\\
&-\frac{1}{\hbar^2}
\sum_{j,k}\alpha_j\alpha_kf(\tau)
\int_0^\tau \mathrm{d}s f(s)\,\Big(
C^{\rm phys}_{jk}(\tau-s)
\big[O^{\rm phys}_{\mathcal S,j},
\widetilde O^{\rm phys}_{\mathcal S,k}(s-\tau)
\sigma^{\rm phys}_{\mathcal S}(\tau)\big]
\nonumber\\
&\hspace{32mm}
+(C_{jk}^{\rm phys})^*(\tau-s)
\big[\sigma^{\rm phys}_{\mathcal S}(\tau)
\widetilde O^{\rm phys}_{\mathcal S,k}(s-\tau),
O^{\rm phys}_{\mathcal S,j}\big]
\Big)
+\mathcal O(\alpha^3)\nonumber,
\label{eq:TCL2relational}
\end{align}
where we have denoted the physical relational density matrix of the system by $\sigma^{\rm phys}_{\S}(\tau)$.
Here, $\widetilde{O}^{\rm phys}_{\S,k}( s-\tau )=\exp(\frac{i(s-\tau)}{\hbar} {H^{\rm phys}_{\S}})O^{\rm phys}_{\S,k}\exp(-\frac{i(s-\tau)}{\hbar}{H^{\rm phys}_{\S}})$ and analogously $\widetilde{O}^{\rm phys}_{\E,j}( s-\tau )$ denote operators in the interaction picture with $H^{\rm phys}_{\E}$, and the physical environmental correlation functions are given by
\begin{equation*}
  C_{jk}^{\rm phys}( \tau-s )=\operatorname{Tr}_{\E}\big(\widetilde{O}^{\rm phys}_{\E,j}(\tau- s )\,
    O^{\rm phys}_{ \E,k}\,\rho^{\mathrm{phys,ref}}_{\E}\big)\,.
\end{equation*}
Similarly, a corresponding Lindblad equation would be relational, would describe evolution in relational time $\tau$, and would involve physical operators and density matrices. Note that in this brief summary we have assumed that the constraint is linear in the clock momentum. If this is not given a priori, then for many models we can rewrite the constraint at the classical level such that it does involve the clock momentum linearly. However, this will modify the explicit form of the physical Hamiltonian, which then, in general, is a function of $H_{\S},H_{\E}$ and $T$. Therefore, reduced phase space quantization can also be applied in the more general case and a relational master equation can be derived. 
In the case of reduced phase space quantization, we solve the constraints completely at the classical level by constructing Dirac observables. 
Although the reference frame, and hence the clock, is treated here as part of the system on which it backreacts, we do not consider its quantum properties and only the system and the environment are quantized.
One of the aims of this work is to extend the formalism of open quantum mechanical systems to situations where one would also like to consider quantum clocks.

\subsection{Non-interacting quantum clocks and relational quantum dynamics}\label{sec:qclocks-rev}
\begin{figure}[!h]
  \centering
    \includegraphics[width=0.5\linewidth]{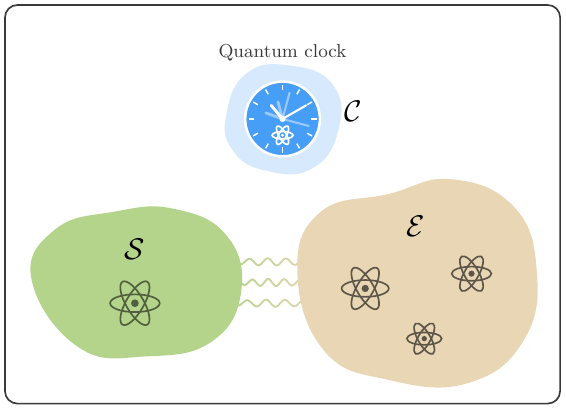}
\caption{An illustration of a relational open quantum system with a non-interacting quantum clock, the quantum clock $\C$ does not couple to $\S$ or $\E$ but is quantized along with the other degrees of freedom.}
\label{fig:nonint-qrf}
\end{figure}

The next generalization of the open quantum system that we want to consider is to treat the clock as quantum as well. As a first step, we briefly review the situation in which the clock does not interact either with the system $\S$ or the environment $\E$, as illustrated in Fig.~\ref{fig:nonint-qrf}. 
This is the case mostly considered in the existing literature. In section~\ref{sec:relational-master-equation} we will present the more general situation with interacting quantum clocks. Since this requires us to generalize some of the methods used for non-interacting clocks, we will briefly review the necessary notions and notation needed for the upcoming sections of the article.

As discussed in subsection \ref{sec:rel-tcl-crfs}, in a reparameterization-invariant theory, the total Hamiltonian is a first-class constraint and there is no external time. This is the problem of time from canonical quantum gravity~\cite{Kuchar:1991qf,Isham:1992ms}. In the context of canonical quantum gravity, the problem of time can be circumvented by introducing a dynamical reference frame and then applying a reduced phase space quantization, see for instance ~\cite{Thiemann:2004wk,Brown:1994py,Giesel:2007wn,Domagala:2010bm,Husain:2011tk,Giesel:2012rb,Giesel:2016gxq}. If one applies a reduced phase space quantization, as discussed in subsection \ref{sec:rel-tcl-crfs}, one, by construction, treats the dynamical reference frame or the clock classically and formulates the models in the perspectival approach ~\cite{Vanrietvelde:2018pgb,Hohn:2018toe,Hoehn:2019fsy,delaHamette:2021oex,AliAhmad:2021adn}.

In order to include the clock as a quantum object in the quantization step, we adopt the perspective-neutral approach to quantum reference frames~\cite{Vanrietvelde:2018pgb,Hohn:2018toe,Hoehn:2019fsy,delaHamette:2021oex,AliAhmad:2021adn} (see also~\cite{Aharonov:1967zza,Aharonov:1984zz,Bartlett:2006tzx,Giacomini:2017zju} for earlier and alternative formulations of quantum reference frames). In this approach a timeless description of a total system is formulated via a constraint and relational descriptions relative to a chosen quantum clock are derived from it.  
 
The perspective neutral approach involves following Dirac quantization ~\cite{Dirac1964,Henneaux:1992ig,Thiemann:2007pyv} wherein we quantize the full kinematical phase space and impose the constraints of the model in the quantum theory. For the setting considered in this work, as before, we divide the non-clock degrees of freedom into a system $\S$ and an environment $\E$.\footnote{What the literature on the Page-Wootters formalism and on quantum clocks~\cite{Page:1983uc,Smith:2019imm,Hoehn:2019fsy} calls ``the system" is therefore the composite $\SE $ here, and the usual system Hamiltonian $H_\S $ and relational state $\rho_\S (\tau)$ correspond to $ {H}_{\SE }$ and $ {\rho}_{\SE }(\tau)$ respectively. This is only a matter of semantics, which we adopt to keep the open quantum system structure visible.}

Quantization of the kinematical phase space yields a factorized kinematical Hilbert space $\mathcal{H}^{\rm kin} = \mathcal{H}_{\C }\otimes\mathcal{H}_{\S }\otimes\mathcal{H}_{\E }$. In most existing works such as~\cite{Page:1983uc,Hoehn:2020epv}, the clock is assumed to not interact with $\SE $. As a result, there is no interaction Hamiltonian and the constraint one starts from is usually of the form
\begin{equation}
    \CH = H_{\C }\otimes\Id_{\SE } + \Id_{\C }\otimes H_{\SE },
    \label{eq:constraint}
\end{equation}
with $H_{\SE } = H_{\S }\otimes\Id_{\E } + \Id_{\S }\otimes H_{\E }$. The classical vanishing of the constraints in the physical sector translates to the requirement that physical states $\rphys$ commute with the constraint operators. For the corresponding flow that the constraint generates, this means that physical states are 
 left invariant, i.e., $e^{-\frac{i}{\hbar}\tau\CH}\rphys \,e^{\frac{i}{\hbar}\tau\CH} = \rphys $ \cite{Dirac1964}.
 
Operators that commute with all constraint operators of the model are Dirac observables in the quantum theory~\cite{Dirac1964,Henneaux:1992ig,Rovelli:1990pi,Rovelli:1990ph,Dittrich:2004cb,Thiemann:2004wk,Chataignier:2019kof,Hoehn:2019fsy}. If the constraints generate gauge transformations, like for instance in general relativity, Dirac observables are gauge invariant quantities and they are defined on the perspective-neutral Hilbert space.

The classical construction of Dirac observables briefly mentioned in subsection~\ref{sec:rel-tcl-crfs} carries over to the quantum theory by using the method of group averaging \cite{Ashtekar:1995zh,Giulini:1998rk,Marolf:2000iq}, in which physical quantities can be obtained by averaging over the flow of $\CH$ requiring $\CH$ to be self-adjoint. A specific realization of group averaging is the $\mathcal{G}$-twirl~\cite{Bartlett:2006tzx,Kitaev:2003zj,Spekkens:2008acy,Loveridge:2017pcv,Castro-Ruiz:2021vnq,Carette:2023wpz}. When applied to the case of one constraint considered here, it has the form
\begin{equation}
    \mathcal{G}(A) = \frac{1}{\mu}\int \mathrm{d}\tau\, e^{-\frac{i}{\hbar}\tau\CH}A\,e^{\frac{i}{\hbar}\tau\CH}.
    \label{eq:Gtwirl}
\end{equation}
After the application of the $\mathcal{G}$-twirl to a given state (density matrix) or operator, they commute with $\CH$ by construction. Here $\mu$ is a normalization constant which is often chosen to be the volume of the parameter set that completely parameterizes the group generated by the constraint and the integral also runs over the same parameter set. We will discuss specific examples in the applications in section~\ref{sec:examples}.

A clock readout is described by a positive operator valued measure (POVM) $\{E_{\C}(\tau)\}$ on $\mathcal{H}_{\C}$~\cite{Holevo1982,BuschGrabowskiLahti1995,Braunstein:1995jb}, which is required to be covariant under the flow of the full constraint $\CH$. For the case of non-interacting clocks the constraint in \eqref{eq:constraint} decomposes into two free operators $H_{\C}$ and $H_{\SE}$ that mutually commute. As a consequence, the corresponding 
unitary operator of $\CH$ factorizes, i.e., $e^{-\frac{i}{\hbar}\tau\CH} = e^{-\frac{i}{\hbar}\tau H_{\C }}\otimes e^{-\frac{i}{\hbar}\tau H_{\SE }}$. Therefore, covariance under $H_{\C }$ is equivalent to covariance under the full constraint $\CH$. 

A clock POVM  can then be built from the eigenstates of the free clock Hamiltonian in two steps. One first forms covariant clock states $\ket{\tau}_\C $ from the eigenstates of $H_{\C }$ and then defines the corresponding clock POVM \cite{Braunstein:1995jb,Smith:2017pwx,Hoehn:2019fsy,Hausmann:2023jpn},
\begin{equation}
    \ket{\tau}_{\C } = \sumint_{\epsilon_n} e^{-\frac{i}{\hbar}\epsilon_n \tau} e^{i\varphi(\epsilon_n)}|\epsilon_n\rangle_{\C },
    \quad
    E_{\C }(\tau) = \ket{\tau}\bra{\tau}_{\C },
    \label{eq:povm}
\end{equation}
that satisfies the following covariance condition~\cite{Holevo1982,BuschGrabowskiLahti1995,Loveridge:2017pcv}
\begin{equation}
e^{-\frac{i}{\hbar}\xi H_{\C }}E_{\C }(\tau)\,e^{\frac{i}{\hbar}\xi H_{\C }} = E_{\C }(\tau+\xi).  
\label{eq:povmCov}
\end{equation}
The clock POVM further satisfies positivity $E_{\C }(\tau)\geq 0$ and completeness $\mu^{-1}\int\mathrm{d}\tau\,E_{\C }(\tau) = \Id_{\C }$.
In the equation above  $\varphi(\epsilon_n)$ is a phase function that can be chosen arbitrarily~\cite{Hoehn:2019fsy}. In the case of non-interacting clocks any choice for $\varphi(\epsilon_n)$ yields the same relational dynamics in the final master equation, thus we work with $\varphi(\epsilon_n)=0$ from now on. Note that the situation is different when we consider interacting clocks for which different choices in the clock POVM in general can yield different relational dynamics. In the qubit clock example in subsection \ref{sec:ex-qubit} physical observables such as the coherence decay rate or the dephasing are independent of these phases. We will comment on this aspect in the applications in section~\ref{sec:examples}. 

For an ideal clock with unbounded spectrum, such as $H_{\C } = p_{\C }$, the clock states are orthogonal and $E_{\C }(\tau)$ constitutes a projection-valued measure (PVM). If the spectrum of $H_\C$ is bounded from below, the overlap is the Fourier transform of a one-sided energy distribution, and by the Paley-Wiener Theorem \cite{PaleyWiener:1934} this forces it to be analytic and square integrable on the real line, properties incompatible with the delta function, so the clock states must overlap regardless of where the spectrum starts. Hence, for such non-ideal clocks the states $\ket{\tau}_{\C }$ overlap and only a covariant POVM is available~\cite{Holevo1982,BuschGrabowskiLahti1995}.

Relational states of the $\SE $ subsystem are obtained by conditioning the physical state on the clock readout. This is the Page-Wootters construction~\cite{Page:1983uc,Wootters:1984wfv} (see also~\cite{Giovannetti:2015qha,Marletto:2016gwv,Hoehn:2019fsy}), which in density operator language reads
\begin{equation}
    \rho_{\SE }(\tau) = \frac{1}{p(\tau)}\mathrm{Tr}_{\C }\big[(E_{\C }(\tau)\otimes\Id_{\SE })\,\rphys \big],
    \qquad
    p(\tau) = \mathrm{Tr}\big[(E_{\C }(\tau)\otimes\Id_{\SE })\,\rphys \big],
    \label{eq:pw}
\end{equation}
where using the cyclicity of the trace, positivity of the density operator and self-adjointness and covariance of the POVM, one can show that $p(\tau)\geq 0$ and $\frac{\mathrm{d}p(\tau)}{\mathrm{d}\tau}=0$. 

In such a non-interacting case, this relational state can be shown to obey the von Neumann equation~\cite{Page:1983uc, Wootters:1984wfv,Giovannetti:2015qha, Hoehn:2019fsy}
\begin{equation}
    \frac{\mathrm{d}}{\mathrm{d}\tau}\rho_{\SE }(\tau) = -\frac{i
}{\hbar}\big[H_{\SE },\rho_{\SE }(\tau)\big].
    \label{eq:vonneumann}
\end{equation}
This shows that relative to a non-interacting clock, $\SE $ evolves unitarily. 

In this section we restricted to the case where no interactions between the quantum clock and $\SE $ are considered, as often done in the existing literature. However, for any realistic clock, such interactions exist~\cite{Zych:2011hu,Pikovski:2013qwa,Castro-Ruiz:2017whw,Smith:2019imm,Paige:2018lrd,Castro-Ruiz:2019nnl} and as discussed in subsection~\ref{sec:rel-tcl-crfs}, already at the classical level, where they are described by adding an interaction term to \eqref{eq:constraint}. 

The aim of the next section is to derive a relational TCL$_2$ master equation with quantum clocks where these interactions are taken into account. We will discuss in detail which of the steps performed so far in the review sections need to be generalized in order to achieve this.

\section{Relational master equation with interacting quantum clocks}\label{sec:relational-master-equation}
\begin{figure}[!h]
  \centering
    \includegraphics[width=0.5\linewidth]{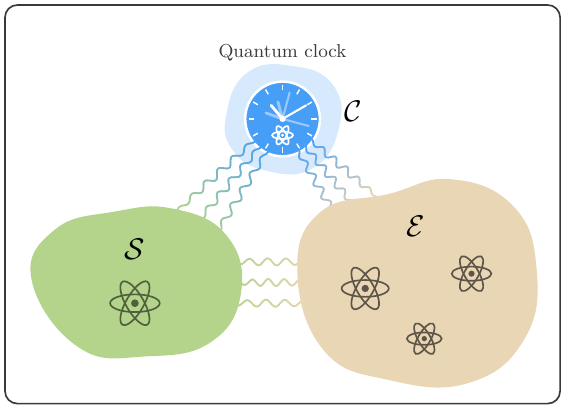}
\caption{An illustration of a relational open quantum system with an interacting quantum clock. The quantum clock $\C$ interacts with both $\S$ and $\E$ and all degrees of freedom are quantized.}
\label{fig:interacting-qrf}
\end{figure}

As discussed in subsection~\ref{sec:qclocks-rev} a quantum clock, in general, will not be isolated from the rest of the degrees of freedom in the theory. In this section we consider quantum clocks that interact with all the non-clock degrees of freedom under consideration, as illustrated in Fig.~\ref{fig:interacting-qrf}.
In general, interactions enter two-fold: On the one hand they change the dynamics of $\rho_{\SE }(\tau)$. On the other hand they also affect the quantum measurement through which we define the clock reading. Consequently in the presence of interactions, the requirement for covariance of the POVM as well as the construction of the POVM itself needs to be generalized. The authors in~\cite{Smith:2017pwx,Hausmann:2023jpn} included interactions along these lines and obtained a modified Schrödinger equation for the dynamics. Their clock POVM is, however, covariant only under $H_{\C }$ and not under the full constraint including the interactions, so the associated  readings do not, in general, account for effects of the interactions on the quantum clock's evolution.
 
The quantum nature of the clock is also expected to cause decoherence of the relational state~\cite{Gambini:2004pe,Pikovski:2013qwa,Castro-Ruiz:2017whw}. Such a decoherence process can be described by a master equation as discussed in subsection~\ref{sec:OQS}. Such equations have so far been obtained largely phenomenologically~\cite{Milburn:1991zkb,Egusquiza:1998zn,Bonifacio:2000xhk,Gambini:2004de,Gambini:2004pe,Gambini:2006yj,Nandi:2026muw}. In each case the decoherence enters through an assumed noise model. To our knowledge no master equation has been derived from a microscopic model of a clock interacting with $\SE $ starting from the perspective-neutral approach.

This is the goal of this section. For this purpose we consider a Hamiltonian constraint with interactions, i.e.,
\begin{equation}
  \CH = H_{\C } + H_{\S } + H_{\E } + \sum_j \alpha_j \Hint{}_{,j},
  \qquad
  \Hint{}_{,j} =  O_{\C ,j} \otimes O_{\S ,j} \otimes O_{\E ,j},
  \label{eq:rme-constraint}
\end{equation}
where the tensor products with the corresponding identity $\Id$ are left implicit in the free contributions, $\alpha_j\in\mathbb{R}$ are coupling constants and the $O_{\C ,j}$, $O_{\S ,j}$, $O_{\E ,j}$ are self-adjoint coupling operators. 
For brevity, we write $H_{\SE } \coloneqq H_{\S } + H_{\E }$ and $\CO \coloneqq H_{\C } + H_{\SE }$ for the free part of the constraint. For notational simplicity, we work with a single interaction term $\Hint = O_{\C } \otimes O_{\SE }$ where $O_{\SE }\coloneqq O_{\S } \otimes O_{\E }$ with a single coupling $\alpha_j=\alpha$. The summation in the interaction term in~\eqref{eq:rme-constraint} can be reintroduced at the end.

\subsection{Clock POVMs for interacting quantum clocks}
\label{sec:GenPOVMInteracting}
In the case of non-interacting clocks as discussed in subsection~\ref{sec:qclocks-rev}, relational time is read off via a positive operator valued measure (POVM) density $E_{\C }(\tau) =  \ket{\tau}\bra{\tau}_{\C }$ in \eqref{eq:povm} that is covariant under the free clock Hamiltonian operator $H_{\C }$. 
As $e^{-\frac{i}{\hbar}\tau\CO}=e^{-\frac{i}{\hbar}\tau H_\C}\otimes e^{-\frac{i}{\hbar}\tau H_{\SE}}$, covariance of $E_\C(\tau)\otimes\Id_{\SE}$ under $\CO$ is equivalent to covariance of $E_\C(\tau)$ under $H_\C$. For $\alpha\neq0$, when interactions are present, the unitary operator $e^{-\frac{i}{\hbar}\tau\CH}$ does not factorize anymore. As a consequence $E_\C(\tau)\otimes\Id_{\SE}$ is no longer covariant under the total dynamics. In this work we do not aim to develop methods for the most general case, but rather consider some constructions that work for a class of models discussed in the literature where so far interactions of the clock have not been taken into account (see for example~\cite{delaHamette:2021oex, Hoehn:2019fsy, Page:1983uc, Wootters:1984wfv}). In future work we plan to generalize the methods developed here to more general situations.

The strategy we will follow here in order to incorporate the interactions in the clock reading is to construct a POVM $E(\tau)$ that is covariant under the full constraint operator $\CH$. Thus, we require
\begin{equation}
  E(\tau+\xi) = e^{-\frac{i}{\hbar}\xi\CH} E(\tau)\: e^{\frac{i}{\hbar}\xi\CH},
  \qquad\text{or equivalently}\qquad
  \frac{\mathrm{d}}{\mathrm{d}\tau} E(\tau) = -\frac{i}{\hbar}\:[\CH, E(\tau)].
  \label{eq:rme-covariance}
\end{equation}
Furthermore, in the special case where we switch off the interactions ($\alpha=0$) we want the generalized POVM $E(\tau)$ to agree with $E_\C(\tau)$, the POVM constructed in the non-interacting case. This allows us to identify the contributions to the master equation that arise  from taking the quantum clock's interactions into account. Moreover, this formulation is convenient in cases where we would like to apply perturbation theory, as for instance done in the derivation of the master equation. 

If we follow the same spirit as in the non-interacting case then clock states from which the POVM is constructed are defined by means of eigenstates of the evolution operator of the clock. Switching on interactions changes this situation and the eigenstates of $H_\C$ used in \eqref{eq:povm} are no longer the eigenstates relevant for the clock's evolution unless we consider $\tau=0$. Therefore, we need something that recalibrates the free clock POVM to the interacting eigenstate basis so that the relational clock reading via $E(\tau)$ is consistently defined with respect to the full dynamics. This can be implemented by what we call a spectral dressing operator $\Omega$ defined as 
\begin{equation}
\Omega\coloneqq \sumint_A e^{i\chi_A}\ket{V_A}\bra{V^{(0)}_A},
\label{eq:SpectralDressing}
\end{equation}
which maps each eigenstate $\ket{V_A^{(0)}}$ of the free part of the constraint $\CO$ to the eigenstate $\ket{V_A}$ of the full constraint $\CH$, where we restrict to models wherein both eigenstates carry the same label, i.e., $\Omega\ket{V^{(0)}_A} = \ket{V_A}$. Here, the phases $\chi_A$ fix the relative phases between eigenstates $\ket{V_A}$, similar to the relative phases in the definition of the clock states in~\eqref{eq:povm}. In the following we choose $\chi_A=0$ for all $A$. In section~\ref{sec:examples}, we will briefly discuss the relevance of these phases for other choices. 

Given the spectral dressing operator, we can then define the generalized POVM given by
\begin{equation}
  E(\tau) = e^{-\frac{i}{\hbar}\tau\CH}\: \Omega \big( E_{\C }(0) \otimes \Id_{\SE } \big)\: \Omega^\dagger\: e^{\frac{i}{\hbar}\tau\CH}.
  \label{eq:rme-povm}
\end{equation}
The spectral dressing $\Omega$ is unitary and acts jointly on $\C$ and $\SE$ Hilbert spaces. As a result, the operator $\Omega(E_{\C }(0)\otimes\Id_{\SE })\Omega^\dagger$, and hence $E(\tau)$ cannot be written, in general, as a tensor product of two operators. 
If the clock is not coupled to the $\SE $ subsystem, then \eqref{eq:rme-povm} reduces to the POVM of subsection~\ref{sec:qclocks-rev}.
One can verify that $E(\tau)$ is positive, complete and fulfills \eqref{eq:rme-covariance} (cf. appendix~\ref{app:povm-properties}).

An alternative, more general strategy, would be to follow the construction in \eqref{eq:povm} with the eigenstates of the full constraint to construct generalized clock states. If one does so the following issues can arise: The full eigenstates might not factorize into clock and remaining degrees of freedom. Directly related to this, in the free theory the tensor product structure ${\cal H}_\C\otimes{\cal H}_\SE$ is preserved by $\CO$, whereas in the interacting case the subspace structure itself changes based on the interaction. Thus defining what we mean by clock degrees of freedom requires further input. Moreover, the fiducial state $\ket{0}_{\C}$ chosen to construct the clock states in the non-interacting case lies in ${\cal H}_\C$, while in the interacting case, such a state needs to be chosen from the full perspective-neutral Hilbert space ${\cal H}^{\rm kin}$. Different choices will generally lead to different POVMs with no canonical choice being preferred. 
The spectral dressing approach sidesteps these by fixing the tensor product structure from the free theory and importing it into the interacting theory via $\Omega$. This is at the cost of requiring $\Omega$ to be well defined. An example of a physical situation in which a spectral dressing operator as written in \eqref{eq:SpectralDressing} will not be well defined is that of level crossings, i.e., if as $\alpha\to 0$, eigenvalues cross in which case the labeling becomes ambiguous. In general, degeneracies in either spectrum can break the one-to-one correspondence assumed above, although the spectral dressing operator can still be used in this form for certain types of degeneracies.

In addition, also if the spectrum changes qualitatively, e.g., bound states appear/disappear between $\CO$ and $\CH$, then the $\Omega$ as defined in \eqref{eq:SpectralDressing} does not exist. We expect that the construction of the spectral dressing operator can be generalized to also be applied when degeneracies or level crossings are present, but we keep that for future work and consider in this article only the simpler and restricted form of $\Omega$ in \eqref{eq:SpectralDressing}.

\subsection{The relational system environmental state for interacting quantum clocks}
\label{sec:RelStateIntClocks}
To obtain relational dynamics with respect to an interacting quantum clock, we generalize the steps done in the construction of subsection~\ref{sec:qclocks-rev}: (i) In the application of the group averaging, for interacting clocks the commutativity of the clock Hamiltonian with the remaining part of the full constraints $\CH$ is no longer given and thus there is no factorization in the ${\cal G}$-twirl. Consequently, even if the kinematical state is initially separable, $\rkin=\rkin_{\mathcal C}\otimes\rkin_{\SE}$ with $\rkin_\SE=\rkin_\S\otimes\rkin_\E$, the $\mathcal G$-twirl can generate correlations in the interacting theory. By contrast, in the non-interacting model, the $\mathcal G$-twirl cannot generate correlations from an initially separable state. This distinction is also relevant for the derivation of the TCL$_2$ master equation, which typically assumes an initial factorized state of the system and environment. (ii) In the non-interacting case relational conditioning is done in the spirit of Page-Wootters. Here we follow the same idea but with a POVM adapted to interacting clocks yielding a generalized Page-Wootters clock conditioning. We then consider the combination of both steps into a single map that yields the relational $\SE$ density matrix from a given initially separable kinematical one $\rkin=\rkin_{\mathcal C}\otimes\rkin_{\SE}$ with $\rkin_\SE=\rkin_\S\otimes\rkin_\E$. 

{\bf $\cal G$-twirl:} Physical states $\rphys$ can be obtained from kinematical ones $\rkin$ by group averaging over the group generated by the full constraint $\CH$, 
\begin{equation*}
  \rphys = \mathcal{G}[\rkin] = \frac{1}{\mu}\int \mathrm{d}s\; e^{-\frac{i}{\hbar}\CH s}\:\rkin\:e^{\frac{i}{\hbar}\CH s},
\end{equation*}
which guarantees $[\CH,\rphys]=0$ and thus implements the gauge-invariance at the quantum level. At the kinematical level we choose an uncorrelated state $\rkin = \rkin_{\C}\otimes\rkin_{\SE}$ with $\rkin_\SE=\rkin_\S\otimes\rkin_\E$. Since $\CH$ contains the interaction term, $\mathcal{G}$ in general correlates clock, system and environment, so that $\rphys$ is no longer of product form.

{\bf Generalized POVM:} For an interacting clock, the interaction affects not only the relational evolution of the total system but, in general, also the reading of the clock because it changes dynamically. Following the spirit of Page-Wootters, for interacting clocks, this should be reflected in the construction of the clock POVM. The POVM $E_{\C}(\tau)$ used in the non-interacting case is covariant only under the free part of the constraint, i.e., $[\Hint,E_{\C}(\tau)\otimes\Id_{\SE}]\neq 0$. Therefore, conditioning by means of $E_\C(\tau)$ on ${\cal H}_\C$ would not be compatible with the flow generated by $\CH$. We therefore condition with the POVM $E(\tau)$ introduced in \eqref{eq:rme-povm}, which is covariant under the full constraint. $E(\tau)$ is constructed in such a way that it reduces to $E_{\C}(\tau)\otimes\Id_{\SE}$ when interactions are switched off, that is  for $\alpha\to 0$. In contrast to previous treatments of interacting clocks, which retain covariance under $H_{\C}$~\cite{Smith:2017pwx,Hausmann:2023jpn}, the clock reading itself formulated by means of $E(\tau)$ is thus adapted to the interaction.

{\bf Generalized Page-Wootters conditioning:} As $E(\tau)$ acts non-trivially on the $\SE$ Hilbert space, a conditioning via $\mathrm{Tr}_{\C}\big(E(\tau)\rphys\big)$ is, in general, neither self-adjoint nor positive, as $E(\tau)$ and $\rphys$ need not commute. We therefore use the symmetric (Lüders-type) update rule, which guarantees a self-adjoint and positive relational $\SE$ state,
\begin{equation}
  \rho_{\SE}(\tau)=\frac{\mathrm{Tr}_{\C}\Big(\sqrt{E(\tau)}\:\mathcal{G}\big[\rkin_{\C}\otimes\rkin_{\SE}\big] \: \sqrt{E(\tau)}\Big)}
{\mathrm{Tr}\Big(\sqrt{E(\tau)}\:\mathcal{G}\big[\rkin_{\C}\otimes\rkin_{\SE}\big]\:\sqrt{E(\tau)}\Big)}
=
p(\tau)^{-1}\: \mathrm{Tr}_{\C }\big(E(\tau)\: \mathcal{G}\big[\rkin_{\C}\otimes\rkin_{\SE}\big] \:E(\tau)\big),
  \label{eq:rme-relstate}
\end{equation}
where the normalization  factor is defined to be $p(\tau) \coloneqq \mathrm{Tr}\big(E(\tau)\:\rphys \:E(\tau)\big)$. In the last step we have used that $E(\tau)^2 = \braket{\tau|\tau}_\C\: E(\tau)$, which follows from unitarity of both $\Omega$ and $e^{-\frac{i}{\hbar}\tau\CH}$, so that $\sqrt{E(\tau)} = E(\tau)/\sqrt{\braket{\tau|\tau}_\C}$. Taking into account the covariance of the POVM $E(\tau)$ and invariance of the physical state under the flow of $\CH$, one obtains $\frac{\mathrm{d}}{\mathrm{d}\tau} p(\tau) = 0$, i.e., $p(\tau)$ is constant in relational time. As a consequence, the normalization will not enter into the final relational master equation because its contributions exactly cancel.
Note here that for $\alpha=0$, the non-interacting case,  this definition of the relational state $\rho_{\SE}(\tau)$ reduces to the usual Page-Wootters one \eqref{eq:pw} from subsection~\ref{sec:qclocks-rev}. 

For our later derivation of the relational master equation, it is convenient to introduce a superoperator $ \Lambda_\tau$ with $\Lambda_\tau:\mathcal T(\mathcal H_{\SE})\longrightarrow\mathcal T(\mathcal H_{\SE})$, where ${\cal T}$ denotes the set of trace-class operators, that maps a kinematical $\SE $ state to a (unnormalized) relational $\SE $ state for a fixed kinematical clock state $\rkin_{\C }$, given by
\begin{equation}
  \Lambda_\tau[X] \coloneqq \mathrm{Tr}_{\C }\Big(E(\tau)\: \mathcal{G}\big[\rkin_{\C } \otimes X\big]\: E(\tau)\Big).
   \label{eq:rme-Lambda}
\end{equation}
Then we can  write the relational state \eqref{eq:rme-relstate} as 
\begin{equation}
\rho_{\SE }(\tau) = p(\tau)^{-1}\Lambda_\tau[\rkin_\SE].  
\label{eq:DefrelStateLamda}
\end{equation}
In the next subsection we will investigate its dynamics with respect to relational time and trace out the environmental degrees of freedom to obtain the relational master equation that we are interested in.

\subsection{Relational master equation with interacting quantum clocks}
\label{sec:DerRelMasterEqn}
The corresponding relational evolution equation for $\rho_\SE$ can be obtained by differentiating \eqref{eq:rme-relstate} with respect to relational time. Taking into account $[\CH,\rphys]=0$, the covariance of the POVM \eqref{eq:rme-covariance}
as well as that the normalization is constant, $\frac{\mathrm{d}}{\mathrm{d}\tau}p(\tau)=0$, yields
\begin{equation}
  \frac{\mathrm{d}}{\mathrm{d}\tau} \rho_{\SE }(\tau)
  = p(\tau)^{-1}\Big( \mathcal{L}_{\SE }\big[\Lambda_\tau[\rkin_{\SE }]\big]
  - \frac{i}{\hbar}\alpha\:\big[O_{\SE },\: \Phi_\tau[\rkin_{\SE }]\big]\Big),
   \label{eq:rme-relEvol}
\end{equation}
where $\mathcal{L}_{\SE }[\ast] \coloneqq -\frac{i}{\hbar}[H_{\SE }, \ast]$ denotes the $\SE $ Liouvillian and we introduced the superoperator $\Phi_\tau:\mathcal T(\mathcal H_{\SE})\longrightarrow\mathcal T(\mathcal H_{\SE})$ defined as
\begin{equation*}
  \Phi_\tau [X] \coloneqq \mathrm{Tr}_{\C }\Big( (O_{\C }\otimes\Id_{\SE })\: E(\tau)\: \mathcal{G}\big[\rkin_{\C } \otimes X\big]\: E(\tau)\Big).
\end{equation*}
The relational evolution equation \eqref{eq:rme-relEvol} still refers to the kinematical state $\rkin_{\SE }$ on the right hand side and is therefore not a closed differential equation for $\rho_\SE(\tau)$. 

Under the condition that the  superoperator $\Lambda_\tau $ can be inverted, we can solve \eqref{eq:DefrelStateLamda} for $\rkin_{\SE }$ and may substitute $\rkin_{\SE } = p(\tau)\:\Lambda_\tau ^{-1}[\rho_{\SE }(\tau)]$.  An analogous invertibility requirement arises in the standard derivation of the time-convolutionless (TCL) master equation, where the invertibility of an operator relating the relevant and irrelevant components of the density operator is required in order to obtain a closed, time-local evolution equation for the reduced state.

Substituting this expression into the evolution equation yields a closed, time-local equation for the conditional system–environment state $\rho_\SE$, where the normalization exactly drops out
\begin{equation}
  \frac{\mathrm{d}}{\mathrm{d}\tau}  \rho_{\SE }(\tau) = \mathcal{K}_\tau [\rho_{\SE }(\tau)],
  \label{eq:rme-kernel}
\end{equation}
with the kernel $\mathcal{K}_\tau $ given by 
\begin{equation}
    \mathcal{K}_\tau [X] \coloneqq  \mathcal{L}_{\SE }[X] - \frac{i}{\hbar}\alpha\:\big[O_{\SE },\: (\Phi_\tau \:\circ\:\Lambda_\tau ^{-1})[X]\big]
  = (\dot\Lambda_\tau  \circ \Lambda_\tau ^{-1})[X].
  \label{eq:Def-kernel-tau}
\end{equation}
Here $\dot\Lambda_\tau $ denotes the derivative of the superoperator $\Lambda_\tau $ with respect to the relational time $\tau$ and $\circ$ denotes the composition of two superoperators. 
Equation~\eqref{eq:rme-kernel} is the exact master equation for the $\SE $ subsystem relative to the clock $\C$. 

As in subsection~\ref{sec:OQS}, we are interested in the dynamics of the system alone under the effective influence of the environment and therefore trace out the environmental degrees of freedom via $\rho_{\S }(\tau) \coloneqq \mathrm{Tr}_{\E }\:\big(\rho_{\SE }(\tau)\big)$.
Writing out the terms, we obtain
\begin{align}
  \frac{\mathrm{d}}{\mathrm{d}\tau} \rho_{\S }(\tau)
  &= -\frac{i}{\hbar}\:\big[H_{\S },\: \rho_{\S }(\tau)\big]\nonumber\\
  &\hspace{1cm}- \frac{i}{\hbar}\alpha\:\Big[O_{\S },\: \mathrm{Tr}_{\CE}\Big( (O_{\C }\otimes\Id_{\S }\otimes O_{\E })\: E(\tau)\: \mathcal{G}\big[\rkin_{\C } \otimes \Lambda_\tau ^{-1}[\rho_{\SE }(\tau)]\big]\: E(\tau) \Big)\Big].
  \label{eq:rme-exact-S}
\end{align}
The evolution equation for $\rho_{\S }(\tau)$ in \eqref{eq:rme-exact-S} is exact but not closed: it requires both the spectral decomposition of $\CH$  for $E(\tau)$ and $\Lambda_\tau^{-1}$ and the knowledge of $\rho_\SE$, rather than $\rho_\S(\tau)$ alone. It can nevertheless be computed exactly when the full spectrum of $\CH$ is known, as for instance in the qubit example in subsection~\ref{sec:ex-qubit}. In general, however, the eigenstates of $\CH$ might not be known in closed form and then we derive instead a perturbative relational master equation by expanding the $\cal G$-twirl, the spectral dressing and the kernel in the coupling parameter $\alpha$ to $O(\alpha^2)$ in analogy to the TCL${}_2$ equation discussed in subsection~\ref{sec:OQS}.  

{\bfseries Perturbative expansion of the relational master equation:} As we are finally interested in a relational TCL${}_2$ master equation with interacting quantum clocks, we need to perform a perturbative expansion of the individual quantities involved in \eqref{eq:rme-exact-S} and collect all terms up to order $O(\alpha^2)$. Since $\Phi_\tau$ appears in the kernel with an explicit factor of $\alpha$, and the dissipator comes from $\Lambda_\tau$ which is itself built from products of first order quantities, expanding $\Omega$ and $E(\tau)$ to first order in $\alpha$ is both necessary and sufficient to obtain the master equation consistently to $O(\alpha^2)$. The second order pieces $\Omega^{(2)}$ and $E^{(2)}(\tau)$ would contribute only at $O(\alpha^3)$ and can be dropped at this truncation order.

For the spectral dressing, up to $O(\alpha)$, we obtain
\begin{equation*}
    \Omega = \Id + \alpha\:\Omega^{(1)} + O(\alpha^2) ,
\end{equation*}
where $\Omega^{(0)}=\Id$ due to completeness of the unperturbed states and as $\ket{V_A}\to\ket{V^{(0)}_A}$ as $\alpha\to 0$. The POVM can be then written up to $O(\alpha)$ as
\begin{equation*}
    E(\tau) = E_{\C }(\tau)\otimes\Id_{\SE } + \alpha\:\big[ \Omega^{(1)}-i\hbar^{-1}\tau W^{(1)}, E_{\C }(\tau)\otimes\Id_{\SE }\big] + O(\alpha^2) ,
\end{equation*}
where $W^{(1)}\coloneqq \sumint_A V_A^{(1)}\ket{V_A^{(0)}}\bra{V_A^{(0)}}$, and $V_A^{(1)}$ is the first-order correction to the eigenvalue of $\CH$. Similarly we can expand the ${\cal G}$-twirl using a Dyson series expansion of the unitaries as $\mathcal{G} = \mathcal{G}^{(0)} + \alpha\:\mathcal{G}^{(1)} + O(\alpha^2)$, where $\mathcal{G}^{(0)}$ is the ${\cal G}$-twirl averaging over the group generated by $\CO$. 

Putting these together, we can write down the corresponding expansions for the superoperators $\Lambda_\tau  = \Lambda^{(0)}_\tau + \alpha\Lambda^{(1)}_\tau + \alpha^2\Lambda^{(2)}_\tau + O(\alpha^3)$ and  $\Phi_\tau =\Phi^{(0)}_\tau +\alpha \Phi^{(1)}_\tau + O(\alpha^2)$, which are needed to construct the perturbations of the kernel ${\cal K}_\tau$ to the appropriate order. The zeroth-order $\Lambda^{(0)}_\tau$ is the superoperator of the non-interacting case.
Assuming that $\Lambda_\tau^{(0)}$ is invertible, we can construct a perturbative inverse of $\Lambda_\tau$ order by order in the coupling strength $\alpha$.

Finally, writing out the full kernel $\mathcal{K}_\tau  = \dot\Lambda_\tau \circ\Lambda_\tau ^{-1}$ of the master equation in orders of $\alpha$ yields
\begin{align*}
  \mathcal{K}^{(0)}_\tau  &= \dot\Lambda^{(0)}_\tau \circ(\Lambda^{(0)}_\tau )^{-1},\\
  \mathcal{K}^{(1)}_\tau  &= \big(\dot\Lambda^{(1)}_\tau  - \mathcal{K}^{(0)}_\tau \circ\Lambda^{(1)}_\tau \big)\circ(\Lambda^{(0)}_\tau )^{-1},\\
  \mathcal{K}^{(2)}_\tau  &= \big(\dot\Lambda^{(2)}_\tau  - \mathcal{K}^{(1)}_\tau \circ\Lambda^{(1)}_\tau  - \mathcal{K}^{(0)}_\tau \circ\Lambda^{(2)}_\tau \big)\circ(\Lambda^{(0)}_\tau )^{-1}.
\end{align*}
Using the covariance condition expanded in $\alpha$, and the invariance of $\rphys$ under $\CH$, one can write, for order $\alpha^k$ with $k\geq 0$:
\begin{equation*}
    \dot\Lambda^{(k)}_\tau [X] = \mathcal{L}_{\SE }\big[\Lambda^{(k)}_\tau[X] \big] - \frac{i}{\hbar}[O_{\SE }, \Phi^{(k-1)}_\tau [\:X\:]],\quad{\rm with}\quad  \Phi_\tau^{(-1)}\coloneqq 0.
\end{equation*}
The label $k-1$ in $\Phi^{(k-1)}$ is derived from the factor $\alpha$ in front of the commutator in the definition of the kernel ${\cal K}_\tau$ in \eqref{eq:Def-kernel-tau}.
Reinserting the appropriate order of $\dot\Lambda^{(k)}_\tau [X]$ into $\mathcal{K}^{(2)}_\tau$, all terms containing $\Lambda^{(2)}_\tau$ cancel, so that the expansion of $\mathcal{K}_\tau$  up to $O(\alpha^2)$ only involves $\Lambda^{(1)}_\tau$ and $\Phi^{(1)}_\tau$.
Using this in the expansion of the kernel above, collecting all the intermediate expansions of the involved quantities, we can compute the expansion of the kernel $\mathcal{K}_\tau  = \mathcal{K}^{(0)}_\tau  + \alpha\:\mathcal{K}^{(1)}_\tau  + \alpha^2\:\mathcal{K}^{(2)}_\tau  + O(\alpha^3)$. Inserting these results into the evolution equation \eqref{eq:rme-exact-S} the relational master equation to second order in the coupling $\alpha$ reads
\begin{align}
  \frac{\mathrm{d}}{\mathrm{d}\tau}  \rho_{\SE }(\tau)
  &= -\frac{i}{\hbar}\:\big[H_{\SE } + \alpha\:g_\C (\tau)\: O_{\SE },\: \rho_{\SE }(\tau)\big] \nonumber\\
  & \hspace{2cm}- \frac{\alpha^2}{\hbar^2} \int_0^\tau \mathrm{d}\xi\: \Big( G_{\C }(\tau,\xi)\: \big[ O_{\SE },\widetilde{O}_{\SE }(\xi - \tau)\:\rho_{\SE }(\tau)\big] + h.c. \Big) \nonumber\\
  & \hspace{2cm}- \frac{i}{\hbar}\alpha^2\: \big[O_{\SE },\: L^{\Omega}_{\SE }(\tau)\:\rho_{\SE }(\tau) + \rho_{\SE }(\tau)\: L^{\Omega\dagger}_{\SE }(\tau)\big] + O(\alpha^3),
  \label{eq:rme-final}
\end{align}
where operators in the $\SE$ interaction picture are given by $\widetilde{O}_{\SE }(\tau) \coloneqq e^{\frac{i}{\hbar}\tau H_{\SE }} O_{\SE } e^{-\frac{i}{\hbar}\tau H_{\SE }}$, and where we have defined the first and second order clock correlation functions as
\begin{align*}
g_\C (\tau)& \coloneqq \frac{\operatorname{Tr}_\C (E_\C (\tau)O_\C )}{\braket{\tau|\tau}},\\
   G_{\C }(\tau,\xi) &\coloneqq C_\C (\tau,\xi) - g_{\C }(\tau) \: g_\C (\xi),\quad\text{with}\quad C_\C (\tau,\xi)=\frac{\mathrm{Tr}_{\C }\big(E_{\C }(\tau)\: O_{\C }\: \widetilde{O}_{\C }(\xi - \tau)\big)}{\braket{\tau|\tau}},
\end{align*}
where operators in the clock's interaction picture have the form $\widetilde O_{\C}(\tau)\coloneqq e^{\frac{i}{\hbar}\tau H_{\C}}O_{\C}e^{-\frac{i}{\hbar}\tau H_{\C}}$ and
\begin{equation*}
    L^{\Omega}_{\SE }(\tau) \coloneqq \frac{\mathrm{Tr}_{\C }\Big( \big(E_{\C }(\tau)\:\mathring{O}_\C(\tau)\otimes\Id_{\SE }\big)\: e^{-\frac{i}{\hbar}\tau\CO}\:\Omega^{(1)}\:e^{\frac{i}{\hbar}\tau\CO} \Big)}{\braket{\tau|\tau}},\quad \mathring{O}_\C(\tau):=O_{\C } - g_{\C }(\tau),
\end{equation*}
where $\mathring{O}_\C(\tau)$ denotes the centered clock coupling operator associated with $O_\C$ obtained by subtracting the mean value $g_\C(\tau)$ of the clock’s coupling operator $O_\C$ relative to the clock readout at relational time $\tau$. Note that $C_\C(\tau,\xi)$ depends on $\tau$ and $\xi$ separately and not only on their difference.

The relational evolution in \eqref{eq:rme-final} describes a von Neumann equation for $\SE$, modified by three  distinct contributions due to the quantum nature of the clock. The first term describes the unitary evolution under the effective Hamiltonian $H_{\SE} + \alpha g_\C(\tau)O_{\SE}$, where the contribution of the quantum clock at first order enters only through its one-point function $g_\C(\tau)$, replacing the full clock coupling operator but its mean value relative to the clock readout. The quantum clock thus acts as a classical time-dependent parameter at leading order, reducing to the standard time-dependent coupling of subsection~\ref{sec:rel-tcl-crfs} for an ideal clock.

The second term is the dissipator sourced by the quantum fluctuations of the clock above its mean, with the correlation functions $G_\C(\tau,\xi) = C_\C(\tau,\xi) - g_\C(\tau)g_\C(\xi)$ playing the role of a noise kernel in exact analogy with the environmental correlation functions discussed in subsection~\ref{sec:OQS}. It vanishes whenever $O_\C$ is diagonal in the clock states, confirming that quantum clock induced decoherence of $\SE$ is a genuinely quantum effect that is absent for ideal clocks.

The third term arises from the spectral dressing $\Omega^{(1)}$ and has no counterpart in the previous treatments of interacting clocks in \cite{Smith:2017pwx,Hausmann:2023jpn}. It encodes the backreaction of the interacting clock on $\SE$ through the modification of the relational time measurement itself. The operator $L^\Omega_{\SE}(\tau)$ depends on the clock's centered coupling operator $\mathring{O}_\C(\tau) = O_\C - g_\C(\tau)$ projected on the dressed eigenstate basis. This captures the mismatch between relational time as read by the free and the interacting clock.

Thus, the three terms encode clock contributions at successive orders in the quantum character of the clock. Both, the second and third terms vanish for ideal clocks, confirming that they genuinely are quantum clock effects. Together, the three terms capture the mean clock behavior, its quantum fluctuations and backreaction of the interacting clock on the relational time readout respectively.

{\bfseries Tracing out the environmental degrees of freedom:} Given the result in \eqref{eq:rme-final}, we can now trace out the environmental degrees of freedom. We take the environmental reference state  $\rho^{\mathrm{ref}}_{\E }$ (cf. subsection~\ref{sec:OQS}) to be stationary, i.e., $[H_{\E }, \rho^{\mathrm{ref}}_{\E }] = 0$. Following the projection-operat approach of open quantum systems~\cite{Breuer:2002}, we  define the relevant projection superoperator $\mathcal{P}X \coloneqq \mathrm{Tr}_{\E }(X)\otimes\rho^{\mathrm{ref}}_{\E }$ and the irrelevant projection superoperator $\mathcal{Q}$ with  $\mathcal{Q} \coloneqq \Id - \mathcal{P}$. Even if we choose an uncorrelated kinematical state $\rkin_{\SE}=\rkin_{\S}\otimes\rho^{\mathrm{ref}}_{\E}$, the resulting relational state $\rho_{\SE}(0)$ is, in general, correlated. As a result, the usual assumption of $\rho_{\SE}(0)$ being separable, which is often used in the derivation of the TCL equation, cannot be imposed here. As a consequence, we get an additional inhomogeneous term in the final relational master equation, which is absent when we specialize to separable states $\rho_{\SE}(0)$.
The final relational master equation for $\rho_\S (\tau)$ up to $O(\alpha^2)$ has the form
\begin{align}
    \frac{\mathrm{d}}{\mathrm{d}\tau} \rho_{\S }(\tau)
  &=  -\frac{i}{\hbar}\:\big[H_{\S } + \alpha\: g_{\C }(\tau)\: g_{\E }\: O_{\S },\: \rho_{\S }(\tau)\big]\nonumber \\
  & \hspace{2cm}- \frac{\alpha^2}{\hbar^2} \int_0^\tau d\xi\: \Big( G_{\CE  }(\tau,\xi)\: \big[ O_{\S },\widetilde{O}_{\S }(\xi - \tau)\:\rho_{\S }(\tau)\big] + h.c.\Big) \nonumber\\
  & \hspace{2cm}- \frac{i}{\hbar}\alpha^2\: \big[O_{\S },\: \ell_{\S }(\tau)\:\rho_{\S }(\tau) + \rho_{\S }(\tau)\: \ell^{\dagger}_{\S }(\tau)\big] \nonumber\\
  & \hspace{2cm}-\frac{i}{\hbar}\alpha\,g_{\C}(\tau)\Big[O_{\S}, {\cal I}_{\SE}(\tau)\Big]\nonumber\\
  & \hspace{2cm}+ O(\alpha^3),
  \label{eq:rme-final-S}
\end{align}
with 
\begin{equation}
    {\cal I}_{\SE}(\tau):=\,e^{-\frac{i}{\hbar}\tau H_{\S}}\Big(\mathrm{Tr}_{\E}\big(\widetilde O_{\E}(\tau)\rho_{\SE}(0)\big)-g_{\E}\,\rho_{\S}(0)\Big)e^{\frac{i}{\hbar}\tau H_{\S}},
    \label{eq:Inhom-rel-me}
\end{equation}
where operators in the $\SE$ interaction picture are given by $\widetilde{O}_{\S }(\tau) \coloneqq e^{\frac{i}{\hbar}\tau H_{\S }} O_{\S } e^{-\frac{i}{\hbar}\tau H_{\S }}$ and $\widetilde{O}_{\E }(\tau)\coloneqq e^{\frac{i}{\hbar}\tau H_{\E }} O_{\E } e^{-\frac{i}{\hbar}\tau H_{\E }}$. Here the last term on the fourth line is the inhomogeneous TCL contribution encoded in ${\cal I}_\SE$ whose explicit form is given in \eqref{eq:Inhom-rel-me}. It accounts for the initial $\S$-$\E$ correlations in the relational state $\rho_\SE(0)$,  see for instance  \cite{Breuer:2002,Fahn:2022zql} for more details about the origin of this term. Crucially, $\mathcal{I}_{\SE}(\tau)$ is determined entirely by the initial relational state $\rho_{\SE}(0)$ and the free system evolution $e^{\pm\frac{i}{\hbar}\tau H_\S}$,  it carries no dependence on the dynamical state $\rho_\S(\tau)$ at later times and can in principle be computed once from the initial data, so that equation~\eqref{eq:rme-final-S} is a closed inhomogeneous linear differential equation for $\rho_\S(\tau)$ alone.

The joint clock-environmental correlation function $G_{\CE}(\tau,\xi)$ is defined as
\begin{equation*}
    G_{\CE  }(\tau,\xi) \coloneqq  C_{\C }(\tau,\xi)\: C_{\E }(\tau-\xi) - g_{\C }(\tau)\: g_{\C }(\xi)\: g_{\E }^2,\label{eq:rme-GCE}
\end{equation*}
where the environmental correlation functions are
\begin{align*}
    g_{\E } &\coloneqq \mathrm{Tr}_{\E }\big(\rho^{\mathrm{ref}}_{\E }\: O_{\E }\big),\\
  C_{\E }(\tau-\xi) &\coloneqq \mathrm{Tr}_{\E }\big(\widetilde{O}_{\E }(\tau-\xi) \: O_{\E }\: \rho^{\mathrm{ref}}_{\E }\big).
\end{align*}
Unlike $C_\C(\tau,\xi)$, the environmental correlation function depends only on $\tau-\xi$, similar to subsection~\ref{sec:OQS}, since the environmental reference state is stationary. Using the covariance of the zeroeth order clock POVM, the clock correlation function can be rewritten as $C_\C(\tau,\xi)=\mathrm{Tr}_\C\big(\widetilde{O}_\C(\tau-\xi)\,O_\C\,E_\C(\xi)\big)/\braket{\tau|\tau}$, where the clock POVM still carries the argument $\xi$. This is because $E_\C(\tau)$ is covariant under $H_\C$ but, unlike $\rho^{\mathrm{ref}}_\E$, does not commute with it.
The third term in \eqref{eq:rme-final-S} contains the  operator $\ell_\S (\tau)$ on ${\cal H}_\S$ which is given by
\begin{align*}
    \ell_{\S }(\tau) &\coloneqq \mathrm{Tr}_{\E }\big( (\Id_{\S }\otimes\rho^{\mathrm{ref}}_{\E }\: O_{\E })\: L^{\Omega}_{\SE }(\tau) \big)\\
    &=\frac{\mathrm{Tr}_{\CE  }\left(
    (E_\C (\tau)\:(\mathring{O}_{\C }(\tau))\otimes \Id_\S \otimes \rho^{\mathrm{ref}}_{\E } O_\E ) \: e^{-\frac{i}{\hbar}\tau\CO} \Omega^{(1)} e^{\frac{i}{\hbar}\tau\CO} 
    \right)}{\braket{\tau|\tau}}.
\end{align*}
The final relational master equation for $\rho_\S(\tau)$ in \eqref{eq:rme-final-S} has the structure of the TCL${}_2$ equation of subsection~\ref{sec:OQS}, but with four modifications due to the quantum clock: 
\begin{enumerate}[label=(\roman*)]
    \item\label{it:first} At first order, the effective Hamiltonian $H_\S+\alpha\:g_\C(\tau)\:g_\E\:O_\S$ becomes time-dependent through $g_\C(\tau)$. This reflects the fact that $\S$ experiences the clock-environment interaction through the mean value of the clock coupling operator. Thus, the quantum clock acts as a classical time-dependent drive with $g_\C(\tau)$ playing the role of a time-dependent coupling strength generated by the clock. The product $g_\C(\tau)g_\E$ shows that this effective drive requires both the clock and environment to have non-vanishing first order correlation functions. 
    \item At second order, the two-point environmental correlation function $C_\E(\tau-\xi)$ of the standard TCL${}_2$ equation is replaced by the joint correlation function $G_{\CE}(\tau,\xi)$. This replacement has a clear physical meaning: Since the clock mediates the $\S$-$\E$ interaction, the noise experienced by $\S$ is not purely environmental but is modulated by the clock fluctuations. The connected structure of $G_\CE(\tau,\xi)$ ensures that only the genuine joint fluctuations of clock and environment above their respective means contribute to the dissipator for $\S$. The mean field contribution already taken into account in \ref{it:first}, is explicitly subtracted. For an ideal clock $G_{\CE}(\tau,\xi)$ is factored as $f(\tau)f(\xi)G_\E(\tau-\xi)$, recovering classically modulated environmental noise without genuine quantum clock contributions to decoherence.
    \item The term containing $\ell_\S(\tau)$ is a correction with no counterpart in the standard TCL${}_2$ equation or in previous treatments of interacting clocks which were based on a POVM covariant only under $H_\C$ \cite{Smith:2017pwx,Hausmann:2023jpn}. It arises from the spectral dressing $\Omega^{(1)}$, which recalibrates the non-interacting clock POVM to the interacting eigenstate basis. Its physical content is that the interacting clock eigenstates differ from the free ones, so that the act of reading off relational time via $E(\tau)$ is itself modified by the interaction. More concretely, $\ell_\S(\tau)$ involves the centered clock coupling operator $\mathring{O}_\C(\tau) = O_\C - g_\C(\tau)$ projected onto the dressed eigenstate basis. Hence, it encodes the degree to which the clock's quantum state is modified by the interaction at the level of the eigenstates rather than just its means.  The resulting correction to the dynamics of $\S$ reflects the fact that eigenstate dressing shifts the effective relational time reading. This term vanishes whenever $O_\C$ is diagonal in the clock eigenstates as, then, $\mathring{O}_\C=0$ and the free and interacting POVMs coincide to first order.
    \item The final term is the inhomogeneous contribution $\mathcal{I}_{\SE}(\tau)$, which is present only when the relational state $\rho_{\SE}(0)$ carries $\S$-$\E$ correlations at $\tau = 0$. These initial correlations are generically generated by the $\cal G$-twirl acting on the kinematical state, even when the latter is separable. Thus $\mathcal{I}_{\SE}(\tau)$ measures their projection onto the centered environment coupling operator $\mathring{O}_\E(\tau) = \tilde{O}_\E(\tau) - g_\E$. Physically, the initial $\S$-$\E$ correlations act as a coherent driving term for the subsequent evolution of $\S$ that cannot be absorbed into the homogeneous kernel. They inject a memory of the initial preparation into the dynamics that decays as the correlations are washed out by the interaction. Under the assumption of a factorizable kinematical state $\rkin = \rkin_\C\otimes\rkin_\S\otimes\rho_\E^{\rm ref}$,  $\mathcal{I}_{\SE}(\tau)$ is of order $O(\alpha)$, making this term consistently $O(\alpha^2)$ overall. Thus the equation~\eqref{eq:rme-final-S} a closed inhomogeneous linear equation for $\rho_\S(\tau)$ alone. 
\end{enumerate}
As discussed above, if $O_\C$ is chosen such that the clock states $\ket{\tau}_\C$ are its eigenstates, then $G_{\C}$ and $\ell_{\S}$ disappear from the equation and~\eqref{eq:rme-final-S} reduces to the TCL${}_2$ equation, including its inhomogeneous term, for a pure system-environment coupling that is time-dependent, $g_{\C}(\tau)\:O_{\S}\otimes O_{\E}$. 
The inhomogeneous term vanishes when $\mathcal{Q}\rho_\SE(0)=0$, i.e., when $\rho_\SE(0)=\rho_\S(0)\otimes\rho^{\mathrm{ref}}_\E$. For ideal clocks, $\rho_\SE(0)$ can be factorized and the inhomogeneous term ${\cal I}_\SE(\tau)$ vanishes. The example in subsection~\ref{sec:ex-ideal} illustrates such a case.

Finally, for completeness, we reintroduce the sum over multiple terms in the interaction Hamiltonian as discussed earlier in \eqref{eq:rme-constraint}, i.e., $\sum_j \alpha_j O_{\C ,j} \otimes O_{\S ,j} \otimes O_{\E ,j}$. Then the relational master equation takes the following form
\begin{align}
    \frac{\mathrm{d}}{\mathrm{d}\tau} \rho_{\S }(\tau)
   &= -\frac{i}{\hbar}\:\Big[H_{\S } + \sum_j \alpha_j\: g_{\C }^j(\tau)\: g_{\E }^j\: O_{\S ,j},\: \rho_{\S }(\tau)\Big] \nonumber\\
  &\hspace{1.5cm} - \sum_{j,k} \frac{\alpha_j \alpha_k}{\hbar^2} \int_0^\tau d\xi\: \Big\{ G^{jk}_{\CE  }(\tau,\xi)\: \big[ O_{\S ,j},\widetilde{O}_{\S ,k}(\xi - \tau)\:\rho_{\S }(\tau)\big] + h.c. \Big\} \nonumber\\
  &\hspace{1.5cm} - \frac{i}{\hbar} \sum_{j,k} \alpha_j \alpha_k\: \big[O_{\S ,j},\: \ell_{\S }^{jk}(\tau)\:\rho_{\S }(\tau) + \rho_{\S }(\tau)\: \ell^{jk\:\dagger}_{\S }(\tau)\big] \nonumber\\
    &\hspace{1.5cm} - \frac{i}{\hbar}\sum_j \alpha_j\: g^j_{\C}(\tau)\Big[O_{\S,j},\, {\cal I}^j_\SE(\tau)\Big] + O(\alpha^3),\label{eq:rme-final-S-multi}
\end{align}
where all the quantities containing the coupling operators get a label, i.e.,
\begin{align*}
    g_{\C }^j(\tau) &\coloneqq \frac{\mathrm{Tr}_{\C }\big(E_{\C }(\tau)\: O_{\C ,j}\big)}{\braket{\tau|\tau}},
  \\
  g_{\E }^j &\coloneqq  \mathrm{Tr}_{\E }\big(\rho^{\mathrm{ref}}_{\E }\: O_{\E,j}\big),
  \\
  C^{jk}_{\C }(\tau,\xi) &\coloneqq  \frac{\mathrm{Tr}_{\C }\big(E_{\C }(\tau)\: O_{\C,j}\: \widetilde{O}_{\C ,k}(\xi - \tau)\big)}{\braket{\tau|\tau}},
  \\
  C^{jk}_{\E }(\tau-\xi) &\coloneqq  \mathrm{Tr}_{\E }\big(\widetilde{O}_{\E ,k}(\tau - \xi)\: O_{\E ,j}\:\rho^{\mathrm{ref}}_{\E } \big),
  \\
  G^{jk}_{\CE  }(\tau,\xi) &\coloneqq  C^{jk}_{\C }(\tau,\xi)\: C^{jk}_{\E }(\tau-\xi) - g_{\C }^j(\tau)\: g_{\C }^k(\xi)\: g_{\E }^j\: g_{\E }^k,
  \\
  \ell^{jk}_{\S }(\tau) &\coloneqq  \frac{\mathrm{Tr}_{\CE  }\Big( \big(E_{\C }(\tau)\:\mathring{O}_{\C ,j}(\tau) \otimes\Id_{\S }\otimes\rho^{\mathrm{ref}}_{\E }\: O_{\E ,j}\big)\: e^{-\frac{i}{\hbar}\tau\CO}\:\Omega^{(1)}_k\:e^{\frac{i}{\hbar}\tau\CO} \Big)}{\braket{\tau|\tau}},
  \\
  {\cal I}^j_\SE(\tau) &\coloneqq e^{-\frac{i}{\hbar}\tau H_\S}\Big(\mathrm{Tr}_{\E}\big(\widetilde{O}_{\E,j}(\tau)\:\rho_{\SE}(0)\big)-g^j_{\E}\:\rho_{\S}(0)\Big)e^{\frac{i}{\hbar}\tau H_\S},
\end{align*}
where the spectral dressing $\Omega^{(1)}_k$ is such that $\Omega^{(1)}=\sum_k\alpha_k\Omega^{(1)}_k$ and the clocks' centered coupling operators are defined as $\mathring{O}_{\C ,j}(\tau)=O_{\C,j}-g^j_{\C}(\tau)$. This form of the relational master equation allows us to also consider different coupling configurations of the quantum clock. For instance, by choosing the identity operator as the clock sector operator in one of the interaction terms, we can construct models in which the clock $\C$ only couples to the system $\S$, while a separate interaction term encodes the coupling between the system and the environment $\SE$, independently of the clock. In such models, the quantum clock and environmental correlation functions contribute additively rather than multiplicatively, with their respective contributions appearing as separated dissipators in the relational master equation.

This finishes our discussion on the derivation of the relational master equation for the system's density matrix $\rho_\S(\tau)$ taking interacting quantum clocks into account, which is the main goal of this work. In the next section~\ref{sec:examples}, we illustrate these results through three examples of increasing complexity, recovering the standard TCL${}_2$ equation in the appropriate limits and identifying in the general case the genuine quantum clock contributions in these applications.

\section{Application of the relational master equation with (interacting) quantum clocks}\label{sec:examples}
In this section, we apply the relational master equation derived in section~\ref{sec:relational-master-equation} to three models. For each example, we first state the constraint and the properties of the operators involved. We then construct the positive operator valued measure (POVM) $E(\tau)$ from \eqref{eq:rme-povm}, and compute the clock correlation functions and derive the relational master equation. In subsection~\ref{sec:ex-nonint} we consider a non-interacting clock as a special case, while in subsection~\ref{sec:ex-ideal} we discuss an ideal clock that couples via its position operator. The third example is presented in subsection~\ref{sec:ex-qubit}, where we investigate a non-ideal qubit clock coupled to a qubit system.

\subsection{Non-interacting quantum clock}
\label{sec:ex-nonint}
In this example, we consider a non-interacting clock and see that the standard master equation for external non-interacting clocks is recovered. The constraint we consider is
\begin{equation*}
  \CH = H_\C + H_\S + H_\E + \alpha\, \Id_\C \otimes O_\S \otimes O_\E.
\end{equation*}
The covariant clock states are $\ket{\tau}_\C$ as in subsection~\ref{sec:qclocks-rev}, while $O_\S$ and $O_\E$ are self-adjoint. This constraint describes the usual setting of open quantum systems, where $\S$ interacts only with $\E$ and the clock serves as an external reference for time. Since $O_\C=\Id_\C$, the constraint takes the form $\CH = H_\C \otimes \Id_{\SE} + \Id_\C \otimes (H_{\SE} + \alpha\, O_{\SE})$.
  
The eigenstates of the free part of the constraint $\CO=H_\C + H_\S + H_\E$ are $\ket{V^{(0)}_A} = \ket{\varepsilon_m^{(0)}}_\C \otimes \ket{e_a^{(0)}}_{\SE}$ with $A\equiv (m,a)$. Since the clock part of $\CH$ is not touched by the interaction, the eigenstates of the full constraint carrying the same label are $\ket{V_A} = \ket{\varepsilon_m^{(0)}}_\C \otimes \ket{e_a}_{\SE}$, where $\ket{e_a}_{\SE}$ denote the eigenstates of $H_{\SE} + \alpha\, O_{\SE}$ with $\ket{e_a}_{\SE}\to\ket{e_a^{(0)}}_{\SE}$ as $\alpha\to 0$. The spectral dressing, therefore, acts trivially on the clock:
\begin{equation*}
  \Omega = \Id_\C \otimes \sumint_a \ket{e_a}\bra{e_a^{(0)}}_{\SE}.
\end{equation*}
As $\sumint_a \ket{e_a}\bra{e_a^{(0)}}_{\SE}$ is unitary on $\mathcal{H}_\SE$, we have $\Omega\big(E_\C(0)\otimes\Id_{\SE}\big)\Omega^\dagger = E_\C(0)\otimes\Id_{\SE}$. Using that $e^{-\frac{i}{\hbar}\tau\CH} = e^{-\frac{i}{\hbar}\tau H_\C}\otimes e^{-\frac{i}{\hbar}\tau(H_{\SE}+\alpha O_{\SE})}$ factorizes, the POVM simplifies to
\begin{equation*}
  E(\tau) = E_\C(\tau) \otimes \Id_{\SE}.
\end{equation*}

With $O_\C = \Id_\C$, the relational state of $\SE$ obeys exactly
\begin{equation*}
  \frac{\mathrm{d}}{\mathrm{d}\tau} \rho_{\SE}(\tau) = -\frac{i}{\hbar}\big[H_{\SE} + \alpha\, O_\SE,\, \rho_{\SE}(\tau)\big],
\end{equation*}
which is simply the von Neumann equation of the interacting $\SE$ subsystem in relational time $\tau$. From \eqref{eq:rme-final-S} one can see that for $O_\C = \Id_\C$, the  clock correlators simplify to $g_\C(\tau)=1$ and $C_\C(\tau,\xi)=1$. Thus $G_{\CE}(\tau,\xi) = C_\E(\tau-\xi) - g_\E^2 = G_\E(\tau-\xi)$. Furthermore, the centered clock coupling operator $\mathring{O}_\C(\tau) = 0$ and thus $\ell_\S(\tau) = 0$. 
The relational master equation, up to $O(\alpha^2)$, can be written as
\begin{align*}
  \frac{\mathrm{d}}{\mathrm{d}\tau} \rho_\S(\tau)
  &= -\frac{i}{\hbar}\big[H_\S + \alpha\, g_\E\, O_\S,\, \rho_\S(\tau)\big]\nonumber\\
  &\hspace{1cm}- \frac{\alpha^2}{\hbar^2} \int_0^\tau d\xi\, \Big\{ G_\E(\tau-\xi)\, \big[ O_\S,\widetilde{O}_\S(\xi-\tau)\,\rho_\S(\tau)\big] + h.c. \Big\}\\
  &\hspace{1cm}-\frac{i}{\hbar}\alpha\Big[O_{\S},\,e^{-\frac{i}{\hbar}\tau H_{\S}}\Big(\mathrm{Tr}_{\E}\big(\widetilde O_{\E}(\tau)\rho_{\SE}(0)\big)-g_{\E}\,\rho_{\S}(0)\Big)e^{\frac{i}{\hbar}\tau H_{\S}}\Big].
\end{align*}
This is the TCL${}_2$ equation of subsection~\ref{sec:OQS} including its inhomogeneous term. All quantum clock contributions vanish as the clock does not interact with $\SE$, and the master equation of non-interacting clocks is recovered as a special case. If $\rho_\SE(0)$ factorizes, the last term vanishes. Note that this result of recovering the TCL${}_2$ equation holds exactly in $\alpha$ not only perturbatively. Since the clock sector is untouched by the interactions, the spectral dressing factorizes exactly and the POVM reduces the one of the non-interacting case to all orders, confirming that quantum clock effects are strictly absent whenever the quantum clock does not couple to $\SE$, regardless of the strength of the $\SE$ interaction itself.

\subsection{Ideal clock with a position-dependent coupling}
\label{sec:ex-ideal}
In this example, we consider an ideal clock that couples to $\SE$ via a function $f$ of its position operator $q_\C$. 
Before we discuss the model, we briefly clarify what we mean by an ideal clock. For this purpose, we consider the overlap of two clock states in \eqref{eq:povm}, which in general is given by~\cite{Hoehn:2019fsy}
\begin{equation}
  \braket{\tau|\tau'}_\C = \chi(\tau-\tau') =
  \begin{cases}
    \displaystyle\int_{\mathrm{spec}(H_\C)} \hspace{-1cm}\mathrm{d}\varepsilon\; e^{\frac{i}{\hbar}\varepsilon(\tau-\tau')}, & \mathrm{spec}(H_\C) \neq \mathbb{R} \text{ continuous},\\[2ex]
    2\pi\hbar\:\delta(\tau-\tau'), & \mathrm{spec}(H_\C) = \mathbb{R},\\[2ex]
    \displaystyle\sum_{\varepsilon\in\mathrm{spec}(H_\C)}\hspace{-0.5cm} e^{\frac{i}{\hbar}\varepsilon(\tau-\tau')}, & \mathrm{spec}(H_\C) \subsetneq\tfrac{2\pi\hbar}{\tau_{\max}}\mathbb{Z},
  \end{cases}
  \label{eq:clock-overlap}
\end{equation}
where $\mathrm{spec}(H_\C)$ denotes the spectrum of $H_\C$, and $\tau_{\max}$ the period of the clock in the case of a discrete clock spectrum. We refer to the clock as \textit{ideal} if $\mathrm{spec}(H_\C)=\mathbb{R}$. In this case, the clock states are orthogonal, i.e., different clock readings are perfectly distinguishable and the clock POVM $E_\C(\tau)$ constitutes a projection-valued measure (PVM). If instead the spectrum of $H_\C$ is bounded from below, e.g., for a finite-dimensional clock, the overlap cannot be a delta function, thus making the clock non-ideal. In addition, the clock coupling operator $O_\C$ needs to be diagonal in the clock states, as already discussed in section~\ref{sec:relational-master-equation}.
For such a position-dependent coupling of an ideal clock, we will show that the relational master equation derived in section~\ref{sec:relational-master-equation} reduces to the standard TCL${}_2$ equation with a time-dependent coupling $\alpha f(\tau)$, where $\tau$ denotes the relational time. The ideal quantum clock is therefore, as expected, operationally indistinguishable from the classical clock when it comes to the dynamics (cf. subsection \ref{sec:rel-tcl-crfs}). It enters the relational master equation of $\S$ only through the classical function $f(\tau)$ with no genuine quantum clock contributions.
 
The total constraint of the model is given by
\begin{equation*}
\CH = c \: p_\C + H_\S + H_\E + \alpha \: f\Big(\frac{q_\C}{\ell}\Big) \otimes O_\S \otimes O_\E .
\end{equation*}
Here $q_\C$ and $p_\C$ are canonical operators with $[q_\C,p_\C]=i\hbar\Id_\C$, where $q_\C$ has dimensions of length and $p_\C$ of momentum. Hence, $H_\C=c\:p_\C$ has dimensions of energy, where $c$ denotes the speed of light, but any other reference velocity could be chosen as well. Further, $f$ is a real-valued function such that $f(q_\C/\ell)$ is self-adjoint, where we introduced a reference length $\ell$ such that the argument of $f$ is dimensionless.
In the following we choose $\ell=c\:t_{\mathrm{ref}}$ with a reference time $t_{\mathrm{ref}}$, which we set to one, thus allowing us to write $f(q_\C/c)$ and, correspondingly, $f(\tau)$ instead of $f(\tau/t_{\mathrm{ref}})$.
As the coupling depends on the clock's position operator $q_\C$, this constraint, as will be seen, corresponds to a time-dependent $\S$-$\E$ interaction whose time dependence is generated by the clock itself.
An often used form of the coupling would be $f(q_\C/c)=q_\C/c$, but we will not further specify the function $f$ in the example, because its main purpose is to compare this quantum clock model to the relational model in subsection \ref{sec:rel-tcl-crfs}, where an interacting classical clock was considered.
For different choices of $f$ the final relational master equation will be different and whether a Markov approximation can be applied will in general depend on such a choice.

The clock states are the eigenstates of the position operator, $q_\C\ket{\tau}_\C = c\tau\ket{\tau}_\C$, with $e^{-\frac{i}{\hbar}\xi c\:p_\C}\ket{\tau}_\C = \ket{\tau+\xi}_\C$, such that $E_\C(\tau) = \ket{\tau}\bra{\tau}_\C$ is covariant under $H_\C = c\:p_\C$. Here, the states are normalized such that $\braket{\tau|\tau'}=2\pi\hbar \delta(\tau-\tau')$ as in \eqref{eq:clock-overlap}. 

Following the steps outlined in section~\ref{sec:relational-master-equation}, we would next need to construct the spectral dressing $\Omega$ and the POVM $E(\tau)$.
In contrast to subsection~\ref{sec:ex-nonint}, the interaction now involves the clock, and since $f(q_\C/c)$ does not commute with $H_\C$, the argument used there, namely that the clock part of the constraint is untouched by the interaction, no longer applies.
However, since $f(q_\C/c)$ is diagonal in the clock states, the flow generated by the full constraint simply shifts the clock position and, as we will see, at the same time evolves $\SE$ with the time-dependent Hamiltonian $H_\SE+\alpha f(q_\C/c)\,O_\SE$, where the clock position takes the role of time.
As discussed in point~(iii) below \eqref{eq:rme-final-S}, for a clock coupling operator that is diagonal in the clock states, the free and the interacting POVMs coincide to first order, and in the present example this even holds exactly.
Indeed, using \eqref{eq:SpectralDressing} with $\chi_A=0$, one finds that $\Omega$ is diagonal in the clock states and reduces to the identity at the clock reading $\tau=0$.
As a consequence, $\Omega$ leaves $E_\C(0)\otimes\Id_\SE$ invariant and the subsequent flow of $\CH$ only shifts the clock reading while acting unitarily on $\SE$, such that the POVM in \eqref{eq:rme-povm} reduces to $E(\tau)=E_\C(\tau)\otimes\Id_\SE$. Note that the reason for this differs from the one in the earlier example in subsection~\ref{sec:ex-nonint}, where $\Omega$ acted trivially on the clock, whereas here it is due to the coupling operator $f(q_\C/c)$ being diagonal in the clock states. Furthermore, as the clock coupling operator simply takes the value $f(\tau)$ at the clock reading $\tau$, the kernel in \eqref{eq:Def-kernel-tau} reduces exactly to the von Neumann equation of $\SE$ with the time-dependent interaction $\alpha f(\tau)\:O_\SE$. Thus, for this model the explicit construction of $\Omega$ and $E(\tau)$ is not needed, and all clock contributions to \eqref{eq:rme-final-S} are captured by the clock correlation functions, which we compute next.
With $\widetilde{O}_\C(\tau) = e^{\frac{i}{\hbar}\tau c\:p_\C} f(q_\C/c)\: e^{-\frac{i}{\hbar}\tau c\:p_\C} = f(q_\C/c + \tau)$, the first and second order clock correlation functions simplify to\footnote{The factors $\braket{\tau|\tau}$ cancel since operators diagonal in $q_\C$ act on $\ket{\tau}_\C$ by multiplication. This cancellation can be made rigorous by restricting the spectrum of $p_\C$ and then taking a limit.}
\begin{equation*}
  g_\C(\tau) = f(\tau),
  \qquad
  C_\C(\tau,\xi) = f(\tau)f(\xi).
\end{equation*}
Hence $G_\C(\tau,\xi) = C_\C(\tau,\xi) - g_\C(\tau)\, g_\C(\xi) = 0$. This highlights the fact that at each clock readout, the coupling operator takes the sharp value $f(\tau)$ and thus has no quantum fluctuations around its mean value. Similarly, $E_\C(\tau)\,\mathring{O}_\C(\tau) = \ket{\tau}\bra{\tau}_\C\big(f(q_\C/c) - f(\tau)\big) = 0$, such that $L^\Omega_\SE(\tau) = 0$ and $\ell_\S(\tau) = 0$. Note that this holds independently of the explicit form of $\Omega^{(1)}$, since in \eqref{eq:rme-final-S} the spectral dressing enters only through $L^\Omega_\SE(\tau)$. The joint clock-environment correlation function thus factorizes,
\begin{equation*}
  G_{\CE}(\tau,\xi) = f(\tau)f(\xi)\,\big(C_\E(\tau-\xi) - g_\E^2\big) = f(\tau)f(\xi)\, G_\E(\tau-\xi).
\end{equation*}
This is the factorized form that was anticipated in point~(ii) below \eqref{eq:rme-final-S}, i.e., the noise experienced by $\S$ is purely environmental and only modulated by the classical function $f(\tau)$. Moreover, for an ideal clock we can always choose $\rho_\SE(0)=\rho_\S(0)\otimes\rho^{\mathrm{ref}}_\E$ without loss of generality, for which the inhomogeneous term in~\eqref{eq:rme-final-S} vanishes. This is possible since for an ideal clock the map from physical states to relational states at $\tau=0$ is invertible.
From \eqref{eq:rme-final-S} one then obtains
\begin{align*}
  \frac{\mathrm{d}}{\mathrm{d}\tau} \rho_\S(\tau)
  &= -\frac{i}{\hbar}\big[H_\S + \alpha\, f(\tau)\, g_\E\, O_\S,\, \rho_\S(\tau)\big]
  \\
  &\hspace{1cm}- \frac{\alpha^2}{\hbar^2} \int_0^\tau d\xi\, f(\tau) f(\xi)\, \Big\{ G_\E(\tau-\xi)\, \big[ O_\S,\widetilde{O}_\S(\xi-\tau)\,\rho_\S(\tau)\big] + h.c. \Big\}.
\end{align*}
This is the TCL${}_2$ equation that one would obtain for a classical clock $T$ that couples with a function $f(T)$ in the classical interaction $\alpha f(\tau)\, {\cal O}_{T,\S}\otimes {\cal O}_{T,\E}$ as discussed in subsection \ref{sec:rel-tcl-crfs}. The reason why we have an agreement at the level of the relational master equation here is because position eigenstates are eigenstates of clock coupling operator function $f(q_\C/c)$ so the clock POVM produces a sharp classical value of the coupling with no quantum fluctuations. Similar to the classical clock in subsection~\ref{sec:rel-tcl-crfs}, here the ideal quantum clock enters through $g_\C(\tau) = f(\tau)$. At first order, the system evolves under the effective time-dependent Hamiltonian $H_\S + \alpha f(\tau) g_\E O_\S$, and in second order the environmental correlation function is weighted by $f(\tau)f(\xi)$. No genuine quantum clock contributions appear, such as decoherence or dissipation, since $O_\C$ is diagonal in the clock states. These kinds of quantum clock effects can, therefore, arise if the clock states are not eigenstates of $O_\C$, which will be the situation in the example discussed in the next subsection \ref{sec:ex-qubit}. The example further confirms the expected result that for the case of an ideal clock even if interactions are present reduced phase space quantization in the perspectival approach, as applied in subsection \ref{sec:rel-tcl-crfs}, and Dirac quantization in the perspective-neutral approach with additional quantum clock conditioning exactly agree.

\subsection{Qubit clock coupled to a qubit system}
\label{sec:ex-qubit}
The previous two examples showed that quantum clock effects are absent whenever the clock either does not interact with $\SE$ (subsection \ref{sec:ex-nonint}) or is ideal in the sense that its coupling operator is diagonal in the clock states (subsection~\ref{sec:ex-ideal}). In the latter case, the clock POVM produces a sharp classical value for the coupling at each readout, so that the two point correlation function factorizes and thus $G_\C(\tau,\xi)$ vanishes identically. Moreover, the POVM covariant under the full constraint reduces to the one covariant under $H_\C$. As a result, the relational master equation reduces to the one obtained for a classical clock, with no clock-induced decoherence or spectral dressing corrections. For genuine quantum corrections to appear, the clock coupling operator $O_\C$ must not be diagonal in the clock states, i.e., the clock states must not be eigenstates of $O_\C$, so that $G_\C(\tau,\xi)$  is non-vanishing. A model that realizes this is a qubit clock coupled to a qubit system without an environment. The constraint we consider for this model is
\begin{equation*}
  \CH = \hbar\omega\, Z_\C + \hbar\omega\, Z_\S + \alpha \,\big( X_\C\otimes X_\S - Y_\C\otimes Y_\S \big) + \beta \,\big( Z_\C\otimes Z_\S + \Id_\C\otimes\Id_\S \big),
\end{equation*}
where $X$, $Y$, $Z$ are the Pauli operators, and $Z\ket{\uparrow} = \ket{\uparrow}$, $Z\ket{\downarrow} = -\ket{\downarrow}$ define the eigenstates of $Z$. $\hbar\omega\, Z_\C$ and $\hbar\omega\, Z_\S$ are the free Hamiltonians of the clock and the system respectively, each describing a two-level system with energy splitting $2\hbar\omega$. The coupling constants $\alpha,\beta\in\mathbb{R}$ have dimensions of energy. 
In terms of the ladder operators $\sigma^\pm = (X \pm iY)/2$, the $\alpha$-interaction term can be rewritten as $2\big(\sigma_\C^+\otimes\sigma_\S^+ + \sigma_\C^-\otimes\sigma_\S^-\big)$, i.e., it creates or destroys excitations in the clock and the system simultaneously. This is analogous to a two-mode squeezing interaction. The $\beta$-interaction is diagonal in the eigenbasis of the free part of the constraint $\CO=\hbar\omega\, Z_\C + \hbar\omega\, Z_\S $. As a result, it only shifts energies and leaves the eigenstates unchanged. 

A related setting was considered in~\cite{Smith:2017pwx}, wherein the Page-Wootters formalism was used with clocks interacting with the system, leading to a time-non-local Schrödinger equation for the conditional state, which was solved perturbatively for several examples of clock-system interactions. However, the clock POVM used in \cite{Smith:2017pwx} is covariant only under $H_\C$, and no master equation is derived.

The bounded spectrum of the qubit clock forces the clock states to overlap, thus yielding a clock that is non-ideal. Moreover, $X_\C$, $Y_\C$ and $Z_\C$ are not diagonal in the clock states. Thus, genuine quantum clock fluctuations contribute to the dynamics of $\S$ through a non-vanishing $G_\C(\tau,\xi)$. As there is no environment, no projection onto $\S$ is needed and the inhomogeneous term of \eqref{eq:rme-final-S} will be absent. 
This example also admits an exact solution, given in appendix~\ref{app:exact-qubit-result}, allowing us to verify the second order expansion and identify the quantum clock contributions also in the exact case.

The eigenstates of the free part of the constraint are $\ket{V_A^{(0)}}\in\{\ket{\uparrow\uparrow},\ket{\downarrow\downarrow},\ket{\uparrow\downarrow},\ket{\downarrow\uparrow}\}$ with free eigenvalues $V_A^{(0)}\in\{2\hbar\omega,-2\hbar\omega,0,0\}$ respectively. Note that the free spectrum is degenerate, since $\ket{\uparrow\downarrow}$ and $\ket{\downarrow\uparrow}$ both have eigenvalue $0$. This does not affect the construction, as both interaction terms annihilate these states.

As both interaction terms annihilate $\ket{\uparrow\downarrow}$ and $\ket{\downarrow\uparrow}$, while the $\alpha$-interaction only couples $\ket{\uparrow\uparrow}$ and $\ket{\downarrow\downarrow}$, the eigenstates $\ket{V_A}$ of the full constraint can be computed to be
\begin{equation*}
  \ket{V_{\uparrow\uparrow}} = \cos\theta\,\ket{\uparrow\uparrow} + \sin\theta\,\ket{\downarrow\downarrow},
  \qquad
  \ket{V_{\downarrow\downarrow}} = \cos\theta\,\ket{\downarrow\downarrow} - \sin\theta\,\ket{\uparrow\uparrow},
  \qquad
  \theta = \frac{1}{2}\arctan\left(\alpha /(\hbar\omega)\right),
\end{equation*}
with eigenvalues $V_{\uparrow\uparrow}=2\beta + 2\zeta $ and $V_{\downarrow\downarrow}=2\beta - 2\zeta $ respectively, where $\zeta  \coloneqq \sqrt{(\hbar\omega)^2+\alpha ^2}$. The other eigenstates are just $\ket{V_{\downarrow\uparrow}}=\ket{\downarrow\uparrow}$ and $\ket{V_{\uparrow\downarrow}}=\ket{\uparrow\downarrow}$ with eigenvalues $0$. Unlike in subsection~\ref{sec:ex-nonint} where the spectral dressing $\Omega$ acted trivially on the clock sector, and unlike in subsection~\ref{sec:ex-ideal} where $\Omega$ was diagonal in the clock states, here $\Omega$ entangles clock and system sectors.
Using the definition of $\Omega$ \eqref{eq:SpectralDressing} with $\chi_A=0$ as done in section~\ref{sec:relational-master-equation}, the spectral dressing for this example is given by
\begin{equation*}
  \Omega = \ket{\uparrow\downarrow}\bra{\uparrow\downarrow} + \ket{\downarrow\uparrow}\bra{\downarrow\uparrow} + \cos\theta\,\big( \ket{\uparrow\uparrow}\bra{\uparrow\uparrow} + \ket{\downarrow\downarrow}\bra{\downarrow\downarrow} \big) + \sin\theta\, \big( \ket{\downarrow\downarrow}\bra{\uparrow\uparrow} - \ket{\uparrow\uparrow}\bra{\downarrow\downarrow} \big),
\end{equation*}
which does not depend on $\beta$. This is because the $\beta$-interaction is diagonal in the free eigenbasis and does not mix eigenstates, but only shifts eigenvalues. Furthermore, the $\alpha$-interaction term only couples $\ket{\uparrow\uparrow}$ and $\ket{\downarrow\downarrow}$. It then follows that the spectral dressing $\Omega$ acts via a rotation by the angle $\theta$ within this subspace and via the identity on $\ket{\uparrow\downarrow}$ and $\ket{\downarrow\uparrow}$ thus having a block structure.

As mentioned in subsection~\ref{sec:qclocks-rev} there exists some freedom in the choice of phases when defining the POVM. For a non-interacting clock a choice of different phases will not affect the relational dynamics as the phases do not enter the evolution equation. In general, we expect this to be different for interacting clocks and thus want to address this question in this example.
For this purpose we consider non-trivial choices for the phases in the definition of the clock states. Following section~\ref{sec:relational-master-equation}, we first construct the clock states $\ket{\tau}_\C$ and the POVM $E_\C(\tau)$ of the free clock as in \eqref{eq:povm}, from which the POVM $E(\tau)$ covariant under the full constraint is then obtained via \eqref{eq:rme-povm}. The clock states and the corresponding POVM density $E_\C(\tau)$ then read
\begin{align*}
  \ket{\tau}_\C &=  e^{-i\omega\tau}e^{i\varphi_\uparrow}\ket{\uparrow}_\C + e^{i\omega\tau}e^{i\varphi_\downarrow}\ket{\downarrow}_\C ,\\
  E_\C(\tau) &= \Id_\C + \cos\left(2\omega\tau+\varphi\right)\, X_\C + \sin\left(2\omega\tau+\varphi\right)\, Y_\C.
\end{align*}
Here we have defined $\varphi\coloneqq \varphi_\downarrow-\varphi_\uparrow$ where $\varphi_\uparrow$ and $\varphi_\downarrow$ denote the phases in \eqref{eq:povm} assigned to the eigenstates $\ket{\uparrow}_\C$ and $\ket{\downarrow}_\C$ of $H_\C$, respectively. 
Here, the period of the clock is $\tau_{\max}=2\pi/\omega$ such that $\frac{\omega}{2\pi}\int_0^{2\pi/\omega}\mathrm{d}\xi\:E_\C(\xi)=\Id_\C$. Thus, the qubit clock is periodic, and can only resolve relational time modulo $\tau_{\max}$. As discussed in~\cite{Chataignier:2024xxx}, relational dynamics relative to periodic clocks comes with additional subtleties, which, for interacting periodic clocks, we leave for future work.
Inserting $\Omega$ and the spectral decomposition of $\CH$ into \eqref{eq:rme-povm}, the POVM becomes
\begin{align*}
  E(\tau) =  \Id &+ e^{-\frac{2i}{\hbar}(\zeta +\beta)\tau}e^{-i\varphi}\,\big( \cos\theta\ket{\uparrow\uparrow} + \sin\theta\ket{\downarrow\downarrow} \big)\bra{\downarrow\uparrow}
  \\
  &+ e^{-\frac{2i}{\hbar}(\zeta -\beta)\tau}\,e^{-i\varphi}\ket{\uparrow\downarrow}\big( \cos\theta\bra{\downarrow\downarrow} - \sin\theta\bra{\uparrow\uparrow} \big) + h.c.\:.
\end{align*}
When $\alpha =\beta =0$, we have $\theta=0$ and $\zeta=\hbar\omega$. Thus, in this case we recover $E_\C(\tau)\otimes\Id_\S$. In contrast to the example in subsection~\ref{sec:ex-ideal}, in this example, $E(\tau)$ does not reduce to $E_\C(\tau)\otimes\Id_\S$ for non-vanishing couplings, i.e., the clock reading itself is modified by the interaction. Since the $\beta$-interaction does not dress the eigenstates, $\beta$ only enters $E(\tau)$ through the frequencies $2(\zeta\pm\beta)/\hbar$, i.e., the interaction shifts the frequencies of the clock, while $\alpha$ rotates the eigenbasis via $\theta$.

Since the full spectrum of $\CH$ is known in this case, the master equation \eqref{eq:rme-exact-S} can be computed exactly in this example. The exact result is given in appendix~\ref{app:exact-qubit-result}. As the exact expression is lengthy, here we present its expansion to second order jointly in $\alpha$ and $\beta$, which allows for a direct comparison with the general result \eqref{eq:rme-final-S-multi}. This second order relational master equation reads
\begin{align*}
  \frac{\mathrm{d}}{\mathrm{d}\tau}\rho_\S(\tau)
  &= -\frac{i}{\hbar}\Big[ H_\S + \sum_j \alpha_j\, g^j_\C(\tau)\, O_{\S,j},\, \rho_\S(\tau) \Big]
  \\
  &\hspace{0.5cm} - \sum_{j,k}\frac{\alpha_j\alpha_k}{\hbar^2} \int_0^\tau d\xi\, \Big\{ G^{jk}_\C(\tau,\xi)\, \big[ O_{\S,j},\widetilde{O}_{\S,k}(\xi-\tau)\,\rho_\S(\tau)\big] + h.c. \Big\}
  \\
  &\hspace{0.5cm} - \frac{i}{\hbar}\sum_j \alpha_j\,\alpha \, \big[O_{\S,j},\, \ell^j_\S(\tau)\,\rho_\S(\tau) + \rho_\S(\tau)\,\ell^{j\,\dagger}_\S(\tau)\big] + O((\alpha,\beta)^3),
\end{align*}
where $O((\alpha,\beta)^3)$ denotes the third order terms in $\alpha $ and $\beta $ jointly. In order to match the form of the general result \eqref{eq:rme-final-S-multi}, we label the four interaction terms in $\CH$ by $X$, $Y$, $Z$ and $\Id$ and identify
\begin{equation*}
\begin{aligned}
  &\alpha_X=\alpha,\quad O_{\C,X}=X_\C,\quad O_{\S,X}=X_\S,
  &\qquad
  &\alpha_Y=\alpha,\quad O_{\C,Y}=Y_\C,\quad O_{\S,Y}=-Y_\S,\\
  &\alpha_Z=\beta,\quad O_{\C,Z}=Z_\C,\quad O_{\S,Z}=Z_\S,
  &
  &\alpha_{\Id}=\beta,\quad O_{\C,\Id}=\Id_\C,\quad O_{\S,\Id}=\Id_\S,
\end{aligned}
\end{equation*}
such that the sums in the following run over these four labels.
The clock correlation functions $g^j_\C$, $G_\C^{jk}$ and the system operators $\ell^j_\S$ depend on the choice of relative phases $\varphi=\varphi_\downarrow-\varphi_\uparrow$ of the clock states. As there is no environment in this example, these follow from the general definitions below \eqref{eq:rme-final-S-multi} with $O_{\E,j}=\Id_\E$, such that $g^j_\E=1$ and $G^{jk}_{\CE}=G^{jk}_\C$. The first order correlation functions $g^j_\C(\tau)$ can be computed to be
\begin{equation*}
  g^X_\C(\tau) = \cos(2\omega\tau +  {\varphi}),
  \qquad
  g^Y_\C(\tau) = \sin(2\omega\tau +  {\varphi}),
  \qquad
  g^Z_\C(\tau) = 0,
  \qquad
  g^{\Id}_\C(\tau) = 1,
\end{equation*}
and the second order correlators $G^{jk}_\C(\tau,\xi) = C^{jk}_\C(\tau,\xi) -g^j_\C(\tau)\:g^k_\C(\xi)$ are
\begin{align*}
  G^{XX}_\C(\tau,\xi) &= \sin(2\omega\tau +  {\varphi})\sin(2\omega\xi +  {\varphi}),
  &
  G^{XY}_\C(\tau,\xi) &= -\sin(2\omega\tau +  {\varphi})\cos(2\omega\xi +  {\varphi}),
  \\
  G^{YX}_\C(\tau,\xi) &= -\cos(2\omega\tau +  {\varphi})\sin(2\omega\xi +  {\varphi}),
  &
  G^{YY}_\C(\tau,\xi) &= \cos(2\omega\tau +  {\varphi})\cos(2\omega\xi +  {\varphi}),
  \\
  G^{XZ}_\C(\tau,\xi) &= -i\sin(2\omega\tau +  {\varphi}),
  &
  G^{YZ}_\C(\tau,\xi) &= i\cos(2\omega\tau +  {\varphi}),
  \\
  G^{ZX}_\C(\tau,\xi) &= i\sin(2\omega\xi +  {\varphi}),
  &
  G^{ZY}_\C(\tau,\xi) &= -i\cos(2\omega\xi +  {\varphi}),
  \\
  G^{ZZ}_\C(\tau,\xi) &= 1, 
  &
  G^{j\Id}_\C(\tau,\xi) &= G^{\Id k}_\C(\tau,\xi) = 0\quad \forall j,k.
\end{align*}
In contrast to the ideal clock example in subsection~\ref{sec:ex-ideal}, the connected clock correlation functions do not vanish as the clock states are not eigenstates of the clock coupling operators. In particular, $G^{ZZ}_\C$ does not vanish although $Z_\C$ commutes with $H_\C$, since the clock states are superpositions of the eigenstates of $H_\C$, such that $Z_\C$ does not take a sharp value at a given clock reading.
The last ingredient is the spectral dressing term. Since the $\beta$-interaction term is diagonal in the free eigenbasis, the spectral dressing $\Omega^{(1)}$ stems entirely from the $\alpha$-interaction term, which is why only $\alpha$ appears as the second coupling in the last line of the master equation above. Finally, using $e^{-\frac{i}{\hbar}\tau\CO}\,\Omega^{(1)}\,e^{\frac{i}{\hbar}\tau\CO} = \tfrac{1}{2\hbar\omega}\big( e^{4i\omega\tau}\ket{\downarrow\downarrow}\bra{\uparrow\uparrow} - e^{-4i\omega\tau}\ket{\uparrow\uparrow}\bra{\downarrow\downarrow} \big)$, the operators $\ell^j_\S$ are
\begin{align*}
    \ell^X_\S(\tau) &= \frac{i\sin(2\omega\tau+\varphi)}{4\hbar\omega}\,\widetilde{X}_\S\big(\sfrac{\varphi}{2\omega}-\tau\big),\\
    \ell^Y_\S(\tau) &= -\frac{i\cos(2\omega\tau+\varphi)}{4\hbar\omega}\,\widetilde{X}_\S\big(\sfrac{\varphi}{2\omega}-\tau\big),\\
    \ell^Z_\S(\tau) &= -\frac{1}{4\hbar\omega}\,\widetilde{X}_\S\big(\sfrac{\varphi}{2\omega}-\tau\big),\\
    \ell^{\Id}_\S(\tau) &= 0.
\end{align*}
Inserting these into the master equation above, performing the integrals over $\xi$ and collecting the terms, the master equation can be written as
\begin{align*}
  \frac{\mathrm{d}}{\mathrm{d}\tau}\rho_\S(\tau)
  &= -\frac{i}{\hbar}\big[ H_\S + \alpha\,\widetilde{X}_\S(\tau+\sfrac{\varphi}{2\omega}),\, \rho_\S(\tau) \big]\\
  &\hspace{1cm} + \frac{\alpha^2}{4\omega\hbar^2}\, \big[\widetilde{Y}_\S(\tau+\sfrac{\varphi}{2\omega}), [\widetilde{X}_\S(\tau+\sfrac{\varphi}{2\omega}), \rho_\S(\tau)]\big]\\
  &\hspace{1cm} 
  - \frac{\beta^2\tau}{\hbar^2}\, \big[Z_\S, [Z_\S, \rho_\S(\tau)]\big]
  \\
  &\hspace{1cm} + \frac{i\alpha\beta\tau}{\hbar^2}\, \big[\widetilde{Y}_\S(\tau+\sfrac{\varphi}{2\omega}), \{Z_\S, \rho_\S(\tau)\}\big]\\
  &\hspace{1cm} 
  + \frac{i\alpha\beta}{4\omega\hbar^2}\, \big[Z_\S, \{\widetilde{X}_\S(\tau+\sfrac{\varphi}{2\omega}), \rho_\S(\tau)\}\big]+ O((\alpha,\beta)^3).
\end{align*}
Here, the $\beta^2\tau$ and $\alpha\beta\tau$ terms grow linearly in $\tau$, since they arise from correlators that do not decay, e.g.\ $G^{ZZ}_\C=1$, whose integral over $\xi$ grows linearly in $\tau$.
In the exact result of appendix~\ref{app:exact-qubit-result}, they correspond to the leading terms of $\tan(2\beta\tau/\hbar)$ and $\sin(2\beta\tau/\hbar)$, respectively.
The second order expansion therefore holds for $\beta\tau/\hbar\ll 1$.
At first order, the clock acts as a classical time-dependent drive, i.e., the system experiences the transverse fields $\widetilde{X}_\S(\tau+\sfrac{\varphi}{2\omega}) = g^X_\C(\tau)\,X_\S - g^Y_\C(\tau)\,Y_\S$ which depends on mean values $g^X_\C(\tau)$ and $g^Y_\C(\tau)$, in analogy to point~(i) below \eqref{eq:rme-final-S}.
At second order, genuine quantum clock effects appear, which were absent for the ideal clock in subsection~\ref{sec:ex-ideal}. The $\beta^2$-term in the master equation describes a dephasing of $\S$ which arises from the quantum fluctuations of $Z_\C$ in the clock states and is absent for an ideal clock. The $\alpha^2$-term combines terms from the clock correlation functions with the spectral dressing contributions $\ell^j_\S$. The terms that depend on $\widetilde{X}_\S(\sfrac{\varphi}{2\omega}-\tau)$ cancel out, thus leaving only the terms depending on the combination $\tau+\sfrac{\varphi}{2\omega}$. 
The exact master equation and its relation to the second order result are discussed in appendix~\ref{app:exact-qubit-result}.
\begin{figure}[!h]
  \includegraphics[width=\textwidth]{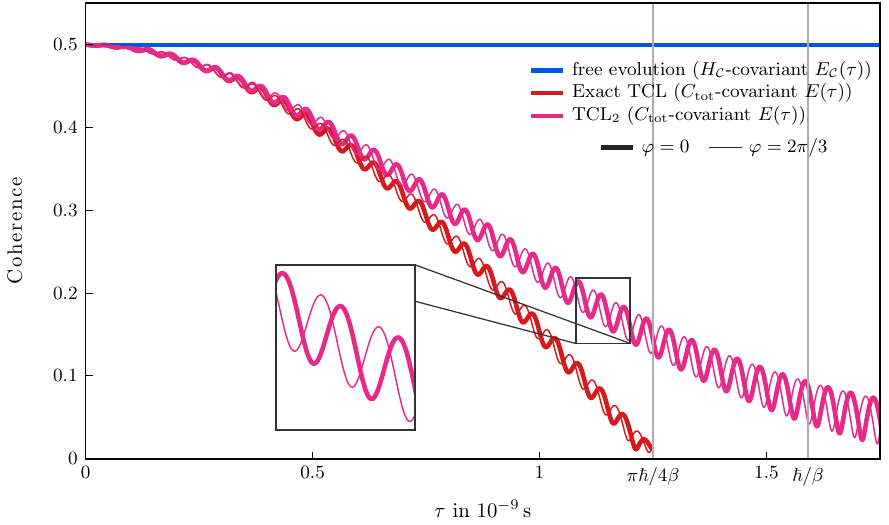}
  \caption{
  Coherence $|\bra{\uparrow}\rho_\S(\tau)\ket{\downarrow}|$ of the relational state of the qubit system for (i) the POVM $E_\C(\tau)$ covariant only under $H_\C$ (free evolution), and for the POVM $E(\tau)$ covariant under $\CH$, evolved with (ii) the exact master equation (exact TCL) and (iii) the second-order master equation (TCL$_2$). Solid lines show the solutions for the relative phase $\varphi=\varphi_\downarrow-\varphi_\uparrow=0$ of the clock states, dotted lines those for $\varphi=2\pi/3$.
  The inset magnifies the marked region of the TCL$_2$ curve, where the phase shifts the fast oscillations while the overall decay is the same for both choices. The free evolution does not depend on $\varphi$. The vertical lines mark $\tau=\pi\hbar/(4\beta)$ where $\cos\big(\tfrac{2\beta\tau}{\hbar}\big)$ first vanishes and the exact master equation ceases to exist because the dynamical map becomes non-invertible (cf. appendix~\ref{app:exact-qubit-result}), and $\tau=\hbar/\beta$, beyond which the second-order master equation is no longer valid. The values chosen for this plot are $\alpha/(\hbar\omega)=0.05$, $\beta/(\hbar\omega)=0.02$, $\omega/2\pi=5\,\mathrm{GHz}$. All curves start from $\rho_\S(0)=\ket{+}\bra{+}$.}
  \label{fig:qubit-coherence}
\end{figure}
To see the effect of these quantum clock contributions on the system, we can compare the relational dynamics obtained from the POVM $E(\tau)$ with those from the POVM $E_\C(\tau)$ covariant only under $H_\C$, looking at the coherence of $\S$, which is affected by the dephasing.

Fig.~\ref{fig:qubit-coherence} shows the coherence of $\S$ relational state as it evolves in relational time $\tau$ in three cases, viz., the evolution equation when the POVM is covariant under the clock Hamiltonian~\cite{Smith:2017pwx}, the exact master equation (cf. appendix~\ref{app:exact-qubit-result}) for when the POVM is constructed as introduced in this work, and the second-order master equation for the same. For the $H_\C$-covariant POVM $E_\C(\tau)$, the system evolves freely and its coherence remains constant as expected for a POVM covariant under $H_\C$~\cite{Smith:2017pwx}. For the $\CH$-covariant POVM $E(\tau)$, in contrast, the coherence decays, which is the dephasing caused by the $\beta$-interaction and thus a genuine quantum clock effect. The exact and the second order results agree for $\beta\tau/\hbar\ll1$, i.e., up to $\tau\approx0.5\times 10^{-9}$~s. At later times, the second order result decays more slowly, since its Gaussian decay $e^{-2\beta^2\tau^2/\hbar^2}$  only reproduces the short-time behavior of the exact factor which is $\cos(2\beta\tau/\hbar)$ (cf. appendix~\ref{app:exact-qubit-result}). The coherence derived from the exact master equation breaks down at $\tau=\pi\hbar/(4\beta)\approx1.25\times 10^{-9}$~s as at this value of $\tau$ the invertibility condition of $\Lambda_\tau$ is not satisfied (cf. appendix~\ref{app:exact-qubit-result}). The underlying reason for this is that at this value of $\tau$ the coherences, involving the factor $\cos(2\beta\tau/\hbar)$, for every initial state vanish and thus the map from  initial relational state to the final relational state is no longer injective, because initial relational states that differ only in its coherences will get mapped to the same relational final state by the superoperator $\Lambda_\tau$. Hence, the superoperator $\Lambda_\tau$ no longer invertible. In the exact solutions this can be directly seen from the divergence of the factor $\tan(2\beta\tau/\hbar)$, where the non-invertibility of $\Lambda_\tau$ carries over to the kernel ${\cal K}_\tau=\dot{\Lambda}_\tau\circ\Lambda_\tau^{-1}$. 
The small oscillations occur at frequency $4\omega$ and do not contribute to the overall decay. 

In the case of non-interacting clocks, such as the example considered in subsection~\ref{sec:ex-nonint}, the choice of phases of $\varphi$ does not show up in the relational dynamics.
With interactions, however, different choices for these phases $\varphi$ yield different relational master equations. 
This deserves a careful interpretation. 
Different choices for the phases $\varphi$ correspond to different reference phases of the clock at the clock reading $\tau=0$.  For the qubit clock, such a choice corresponds to the direction of the Bloch vector on the Bloch sphere at $\tau=0$. For $\varphi=0$ it points along the $X$-direction, whereas for non-vanishing $\varphi$, it is rotated by the angle $\varphi$ in the $XY$-plane. 
Physical observables, e.g., dephasing rates or coherence decay rates, are independent of the choice of phases $\varphi$. In the interacting case, the choice of phases $\varphi$ thus represents a residual freedom in defining the clock reading. In the non-interacting case, a change of $\varphi$ is merely a shift of the time origin, which does not show up in the von Neumann equation generated by $H_\SE$ alone, where the latter does not depend on the choice of $\varphi$.

Similar to $\varphi$, the phases $\chi_A$  appear in the definition of the spectral dressing operator \eqref{eq:SpectralDressing}. These phases modify how the non-interacting and interacting theory are related. Choosing different phases $\chi_A$ transforms the POVM $E(\tau)$ by unitary operators $U_{\chi}=\sum_A e^{i\chi_A}\Pi_A$, where $\Pi_A$ denotes the projector onto the eigenspaces of $\CH$ having eigenvalues $V_A$. For non-degenerate spectra, $\rphys$ is diagonal in the eigenstates of $\CH$. Since $U_\chi$ only multiplies each eigenspace by an overall phase, $U_\chi\rphys U_\chi^\dagger=\rphys$. While the physical state is invariant under $U_\chi$, the relational state is in general not, since $U_\chi$ acts jointly on combined clock-$\SE$ Hilbert space and thus
cannot be removed from the partial trace $\mathrm{Tr}_\C\big(U_\chi E(\tau)\rphys E(\tau)U_\chi^\dagger\big)$. Different choices of $\chi_A$ therefore correspond to different readings of
the interacting clock, analogous to the phases $\varphi$. A more systematic investigation of the role of these phases is planned for future work.

Additionally, in the context of the qubit example we want to address the question whether choosing just the POVM $E_\C(\tau)$, which is covariant only under $H_\C$, or $E(\tau)$, which is covariant under the full constraint, corresponds to just two different choices of viable, but different, clock readings or if there exists some reason for the POVM $E(\tau)$, which is adapted to the interactions, to be preferred. 

For this purpose we consider how a gauge transformation of the interacting model acts on the two different choices of POVMs. By construction, physical states are invariant under this gauge transformation, i.e., $e^{-\frac{i}{\hbar}\xi\CH}\rphys\,e^{\frac{i}{\hbar}\xi\CH}=\rphys$, for any real gauge parameter $\xi$. Applying such a transformation directly to the non-interacting clock POVM yields
\begin{equation*}
    E^{(\xi)}_\C(\tau)=e^{-\frac{i}{\hbar}\xi\CH}\big(E_\C(\tau)\otimes\Id_\S\big)e^{\frac{i}{\hbar}\xi\CH}=E_\C(\tau+\xi)\otimes\Id_\S-\frac{i\xi}{\hbar}\big[\Hint,E_\C(\tau)\otimes\Id_\S\big]+\mathcal{O}(\xi^2),
\end{equation*}
which in the case of the qubit clock example takes the form
\begin{align*}
E^{(\xi)}_\C(\tau)&=E_\C(\tau+\xi)\otimes\Id_\S
+\frac{\xi}{\hbar}\Big[\alpha\,Z_\C\otimes\big(\sin(2\omega\tau+\varphi)X_\S+\cos(2\omega\tau+\varphi)Y_\S\big)\\
&\hspace{3cm}+\beta\,\big(\cos(2\omega\tau+\varphi)Y_\C-\sin(2\omega\tau+\varphi)X_\C\big)\otimes Z_\S\Big]+\mathcal{O}(\xi^2).
\end{align*}
We realize that the gauge-transformed $E_\C^{(\xi)}(\tau)$ can no longer be written as $E_\C(\tau+\xi)\otimes\Id_\SE$, where the effect of the gauge transformation would just be a relabeling of relational time. Instead we get an additional correction term which is linear in the coupling constant and depends explicitly on the gauge parameter $\xi$. In contrast, for the $\CH$-covariant POVM $E(\tau)$ the same gauge transformation only relabels the clock reading $E(\tau)\mapsto E(\tau+\xi)$.  

Next we want to compare the two relational states that we obtain when conditioning using \eqref{eq:rme-relstate} either with respect to the gauge-transformed $E^{(\xi)}_\C(\tau)$ or $E(\tau)$. For the $\CH$-covariant POVM $E(\tau)$ the relational state satisfies the same master equation as the untransformed state 
\begin{equation*}
\frac{\mathrm{d}}{\mathrm{d}\tau}\rho_\S(\tau+\xi)=\mathcal{K}_{\tau+\xi}\,\rho_\S(\tau+\xi),
\end{equation*}
just with $\tau$ replaced by $\tau+\xi$ everywhere, so the master equation describes the same physics and fixing a gauge just corresponds to fixing the origin of relational time.
However, if we condition with respect to the gauge transformed POVM $E^{(\xi)}_\C(\tau)$, then the resulting relational state $\rho^{(\xi)}_\S(\tau)$ obeys the relational master equation
\begin{equation*}
\frac{\mathrm{d}}{\mathrm{d}\tau}\rho^{(\xi)}_\S(\tau+\xi)=\mathcal{K}^{(\xi)}_{\tau+\xi}\,\rho^{(\xi)}_\S(\tau+\xi).
\end{equation*}
Here the relational master equation cannot be brought into the form of the untransformed relational state by a simple relabeling of relational time. To begin with, this is because $\rho^{(\xi)}_\S(\tau+\xi)$ contains additional terms that depend on the gauge parameter $\xi$ in addition to the shifted argument. Moreover, the kernel $\mathcal{K}^{(\xi)}_{\tau+\xi}$ also has additional terms depending on the gauge parameter. Therefore, we see that in general, physical predictions obtained when using a clock conditioning with $E_\C(\tau)$ as the POVM can depend on the chosen gauge, when dealing with interacting clocks. This is something we would like to avoid in general.

\begin{figure}[!h]
\centering
\includegraphics[width=0.98\textwidth]{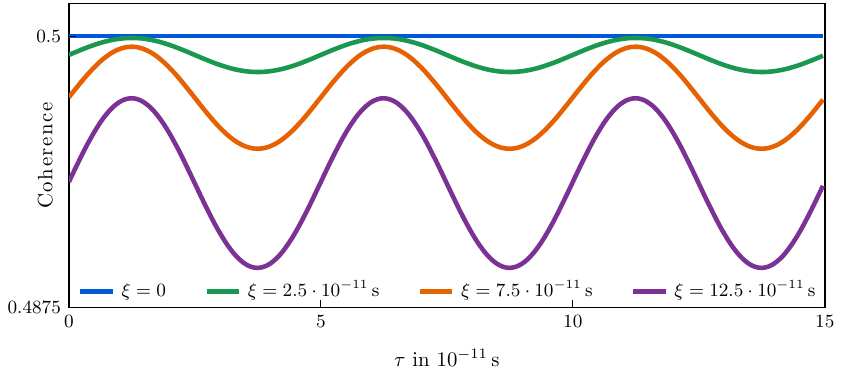}
\caption{Comparison of the evolution of the coherences of $|\bra{\uparrow}\rho^{(\xi)}_\S(\tau)\ket{\downarrow}|$ of the gauge transformed relational state of the qubit system which was defined using the POVM $E_\C(\tau)$ that is covariant only under $H_\C$. The plot includes curves for the values of the gauge parameter $\xi=0$, $2.5$, $7.5$ and $12.5\cdot10^{-11}\,\mathrm{s}$. All curves correspond to the same physical state $\frac{1}{\sqrt{2}}(\ket{\uparrow\downarrow}+\ket{\downarrow\uparrow})$, which yields $\rho_\S(0)=\ket{+}\bra{+}$ for $\xi=0$. For $\xi=0$ the coherence is constant. For $\xi\neq0$ it drops to a $\xi$-dependent level, approximately $\frac12|\cos(2\beta\xi/\hbar)|$, and shows a small oscillation with frequency $4\omega$. The vertical lines mark $\tau=\pi\hbar/(4\beta)$ and $\tau=\hbar/\beta$ as in Fig.~\ref{fig:qubit-coherence}. The values chosen for this plot are $\alpha/(\hbar\omega)=0.05$, $\beta/(\hbar\omega)=0.02$, $\omega/2\pi=5\,\mathrm{GHz}$ and $\varphi=0$.}
\label{fig:qubit-gauge}
\end{figure}

To illustrate this for the qubit example, Fig.~\ref{fig:qubit-gauge} shows the coherence of the relational state obtained by conditioning on $E_\C^{(\xi)}(\tau)$ for different values of the gauge parameter $\xi$.
All curves are computed exactly for the same physical state $\frac{1}{\sqrt{2}}\big(\ket{\uparrow\downarrow}+\ket{\downarrow\uparrow}\big)$, which yields $\rho_\S(0)=\ket{+}\bra{+}$ for $\xi=0$.
For $\xi=0$ the coherence stays constant at $1/2$. For $\xi\neq0$ it drops to a lower level, and in addition acquires a small oscillation in $\tau$. The same physical state thus appears with constant coherence in one gauge, but shifted by a constant value and time dependent in another. As physical quantities, e.g., the evolution of coherence should not depend on gauge artifacts, this example already demonstrates that using the POVM $E_\C(\tau)$, that is not adapted to the interactions of the model, is problematic. Note that for neither choice of gauge we would observe the non-trivial decay rate obtained when we define the relational state with respect to $E(\tau)$. This is because, by construction, $E_\C(\tau)$ has no access to the richer physics encoded in $E(\tau)$, which is, apart from the inconvenient gauge dependence of $E_\C(\tau)$, a further reason why we would prefer to use $E(\tau)$ for interacting quantum clocks.

We finally comment on why such a gauge-dependence of $E_\C(\tau)$ was not already discussed in \cite{Smith:2017pwx}. A difference to our situation here is that in \cite{Smith:2017pwx} they derive a modified (relational) Schrödinger equation and considered state vectors only. 
The construction of \cite{Smith:2017pwx} is internally consistent. The conditional state in \cite{Smith:2017pwx} is defined by projecting the physical state onto a fixed clock state. 
In the examples of \cite{Smith:2017pwx}, the POVM constructed in the current work $E(\tau)$ coincides with the clock POVM $E_\C(\tau)$ of the non-interacting case when we restrict ourselves to matrix elements in physical states. In the qubit example, this is because $\Hint$ annihilates the physical state vectors. In the ideal clock example 4.1 in \cite{Smith:2017pwx}, this is because $\Hint$ commutes with $E_\C(\tau)$ thus meaning that a gauge transformation simply shifts the clock reading. In the gravitational example 4.2 with the ideal clock in \cite{Smith:2017pwx}, $\Hint$ commutes with $\CO$ and thus $\Omega=\Id$ and the form of the constraint implies that the clock evolution acts like a system evolution, which commutes with  $E_\C(\tau)\otimes\Id_\S$.
In the more general case, where this is not necessarily given, requiring that the master equation for the gauge-transformed relational state describe the same physics leads to the $\CH$-covariant POVM $E(\tau)$ which is constructed in this work.
Moreover, in the qubit example, the interaction annihilates the physical state vectors, which is why it has no effect on the conditional state in their case where they consider matrix elements with respect to physical states. In contrast, the conditioned operator $E_\C(\tau)\rphys E_\C(\tau)$ used here has support
outside the kernel of $\CH$, where $H_{\rm int}$ and hence its unitaries act non-trivally. Hence, the $\xi$-dependence shown in Fig.~\ref{fig:qubit-gauge} arises even for these states. Thus, the construction used in \cite{Smith:2017pwx} is internally consistent as one does not see the gauge artifacts in examples therein.
  
Note that the conclusion that in general physical quantities can depend on the chosen gauge if one uses a non-covariant POVM in the case of interacting quantum clocks, is not restricted to the example considered here but just follows from the way how the corresponding master equations of the gauge-transformed relational state  defined by means of a non-covariant POVM look like. In very special cases, such as in \cite{Smith:2017pwx} the correction terms might all vanish so that no gauge-dependence survives in the end, but generically this will not be the case.

Finally, we can also comment on the non-Markovianity of the relational quantum dynamics in this example. There is no unique notion of (non-) Markovianity for quantum dynamics, and several inequivalent definitions are used in the literature~\cite{Plenio:2014nyj,Breuer:2015zlm}. One such notion underlies the derivation of the Lindblad equation mentioned in subsection~\ref{sec:OQS}. Therein, a Markov approximation is performed, in which the upper limit of the integral in~\eqref{eq:TCL2} is sent to infinity, also known as the (second) Markov limit. This is only justified if the correlation functions decay quickly compared to the time scale of $\S$, such that the error made in this step is small. In the qubit example, this is not the case. The clock correlation functions $G^{jk}_\C(\tau,\xi)$ are either constant or oscillating and do not decay. Hence, a Markov approximation cannot be performed for this model. Another notion of Markovianity that is commonly used is that of  CP-divisibility~\cite{Rivas:2010pzm, Plenio:2014nyj,Hall:2014oiy}. The dynamics is CP-divisible if the evolution between any two times is given by a completely positive trace preserving map. This condition is easily understood once the time-local master equation is written in the so-called \textit{Kossakowski form}~\cite{Gorini:1975nb,Hall:2014oiy} (see appendix~\ref{app:exact-qubit-kossakowski}). This form consists of a unitary part and a dissipator, which is fixed by a Hermitian and, in general, time-dependent Kossakowski matrix $\Gamma(\tau)$. The eigenvalues of $\Gamma(\tau)$ are the decoherence rates. The dynamics is said to be CP-divisible if and only if all decoherence rates are non-negative~\cite{Hall:2014oiy}. The Kossakowski form of the exact master equation for the qubit example is given in appendix~\ref{app:exact-qubit-kossakowski}. Also here one can see that one of the two non-zero eigenvalues of the Kossakowski matrix is positive while the other one is negative at all $\tau$. Hence, the relational dynamics of $\S$ is not CP-divisible and, therefore, non-Markovian even in this sense. This is a consequence of the clock being quantum and non-ideal. The clock is a qubit whose correlation functions do not decay. As a result, it `stores' the information it receives from $\S$ and partially returns it to the system. This agrees with the analysis in~\cite{Nandi:2026muw}, where a finite clock resolution was found to induce non-Markovian reduced dynamics, while here the non-Markovianity follows from a microscopic model.

\section{Conclusion and Outlook}
\label{sec:Conclusion}
The main result of this work is a relational master equation with interacting quantum clocks, derived from microscopic models in the context of quantum mechanics, which we have presented in detail in section~\ref{sec:relational-master-equation}. To achieve this we extended the methods in the existing literature that mostly focus on closed systems and non-interacting clocks~\cite{Page:1983uc,Wootters:1984wfv,Hoehn:2019fsy,Hoehn:2020epv,Vanrietvelde:2018pgb,delaHamette:2021oex}. Particularly in the framework of open quantum systems, it is rather restrictive to work with non-interacting clocks. 
Moreover, interactions of the clock do not only modify the dynamics of system and environment but also influence the clock reading itself. To take this into account we generalized the construction of the clock POVM and, with it, also the usual Page-Wootters conditioning.
The strategy we followed in this work here is to first construct a POVM that is covariant under the full constraint and not only the free clock Hamiltonian. This POVM involves a spectral dressing operator which relates the free and interacting theory. This allowed us to fix the tensor product structure in the free theory and import it into the interacting theory. When interactions are switched off, the POVM for the interacting clock agrees with the one of the non-interacting clock. 
With a generalized Page-Wootters conditioning adapted to the interactions present in the model, combined with ${\cal G}$-twirl, we defined the physical relational state of the system-environment and derived its evolution with respect to relational time.
To obtain a closed differential equation for the relational state, we needed to require an invertibility condition on a superoperator that maps a kinematical system-environmental state to its corresponding physical relational state. This is similar to invertibility conditions that are present in the usual derivation of the master equation in the projection operator formalism. As the next step, we then traced out the environmental degrees of freedom to obtain the relational master equation for the system of interest.

The final relational master equation for the system's density matrix  has the structure of the TCL${}_2$ equation of subsection~\ref{sec:OQS}, but with four modifications due to the quantum clock. At first order of the coupling constant, the quantum clock acts like a classical drive. Herein, the one-point correlation function of the clock plays the role of a coupling strength that depends on the relational time. At second order the usual environmental correlation functions are replaced by the joint correlation functions of the clock and environment. Since the clock mediates the system-environment interaction, the noise experienced by the system is not purely environmental but is modulated by the clock fluctuations.
Hence, whether the dynamics of the system is Markovian, depends not only on the microscopic properties of the environment, but also on those of the quantum clock.
Additionally, at second order, we obtained a term arising from the spectral dressing operator. This term has no counterpart in the standard TCL${}_2$ master equation or in previous treatments of interacting clocks which were based on a POVM covariant only under the free clock Hamiltonian. The spectral dressing operator recalibrates the non-interacting clock POVM to the interacting eigenstate basis. Physically, this means that the interacting clock eigenstates differ from the free ones, so that the act of reading off relational time via the POVM in the interacting case is itself modified by the interaction. Furthermore, the relational master equation for the system involves an inhomogeneous contribution, which is present only when the relational state carries system-environment correlations at the initial time. Such initial correlations are generically generated by the $\cal G$-twirl acting on the kinematical state, even if the kinematical state is separable. 
Which of these contributions appear, and in which form, depends on the specific microscopic model that encodes the interactions between the system, environment and quantum clock.

As first applications of the relational master equation we investigated three examples in section~\ref{sec:examples}. In the first example, we considered a non-interacting clock to check whether we get the expected results. As expected, the clock then behaves like an external classical clock and we recover the standard TCL${}_2$ master equation. Next, we considered an ideal clock that interacts with the system and environment. Our results show that in this case, as expected, no genuine quantum clock corrections are present. Moreover, we see that we obtain the same final relational master equation regardless of whether we use reduced phase space quantization or whether we use Dirac quantization in the perspective-neutral approach and then go to the perspective of the quantum clock. Finally, as an example of a non-ideal interacting quantum clock, we analyzed a qubit clock coupled to a qubit system with no environment present. 
In this example, all quantum clock contributions are present in the relational master equation for the system, apart from the inhomogeneous contribution since there is no environment to trace out in this example. For this model, the relational master equation can also be solved exactly, thus further allowing us to compare the perturbative solution with the exact one.

Although the qubit clock example is simple and manageable, it already gives some insights into the physical relevance of these quantum clock effects. 
For this purpose, we compared the coherences of the system when using two different POVMs, namely the POVM covariant under the full constraint and the POVM of the non-interacting case which is covariant only under the free clock Hamiltonian. These correspond to two different readings of the relational time.
In the non-interacting case all quantum clock correlation functions are computed using the non-interacting POVM, whereas in the other case the POVM adapted to the interactions is used. Using the POVM for the non-interacting case in this example, the system evolves freely and we obtain constant coherences. In contrast using the fully adapted POVM that is sensitive to the interaction, the coherences decay in relational time. This decay is caused by the additional modifications to the master equation, which encode backreaction of the quantum clock onto the system and the adapted reading of the relational time using the generalized Page-Wootters conditioning. 
Furthermore, the example was useful to better understand the role of possible phases that can be included in the definition of the POVM and the spectral dressing operator. 
For non-interacting clocks, this choice of phases in the POVM has no influence on the final relational dynamics.
However, for interacting clocks, these phases in general enter the final master equation. 
In the qubit example, choosing different phases in the POVM has a clear physical interpretation: it corresponds to different reference phases at relational time $\tau=0$, which is a residual freedom defining the clock reading in the interacting model. For this example physical quantities such as the decoherence rates or dephasing rates do not depend on this choice of phase.
Similarly, the phases in the spectral dressing operator leave the physical state invariant but in general enter into the relational state, so that they too specifiy the reading of the interacting clock. A more systematic investigation of the role and relevance of these phases is planned in future work.

We further investigated whether a POVM covariant only under the free clock Hamiltonian, rather than the full constraint, can always be used consistently in the presence of interactions, or whether this can be problematic in general. As we showed in subsection~\ref{sec:ex-qubit}, if we perform a gauge transformation on a relational state conditioned on a POVM that is not fully covariant, the corresponding relational master equation does not describe the same physics. This is because, in general, it will depend on the gauge. We demonstrated this in the example of the qubit clock by computing the evolution of gauge-transformed relational state's coherence. For the POVM that is not adapted to the interactions, the evolution of the coherence is constant in one gauge but oscillates in another, thereby leading to gauge artifacts that are unphysical. If we instead use the POVM that is adapted to the interactions, no such gauge-dependence is present. As our results show, we can explicitly compute these gauge-dependent corrections for a given model and it turns out that for the models considered in \cite{Smith:2017pwx} these vanish. This is because, for the examples considered in \cite{Smith:2017pwx}, the POVM constructed in the current work in section~\ref{sec:relational-master-equation}, which is adapted to the interactions, coincides with the clock POVM of the non-interacting case, when we restrict ourselves to the physical states.

Finally, we comment again on the central result of~\cite{Nandi:2026muw}, namely that exact Markovianity is not fundamental once time is defined relationally through a quantum clock with finite resolution.
In \cite{Nandi:2026muw}, the reduced dynamics is generically non-Markovian, and a Markovian description emerges only after a relational coarse-graining, provided the clock resolution scale is large compared to the correlation time of the environment. Instead, in this work, although we consider quantum mechanical rather than relativistic field theory models, we encode the quantum clock at the microscopic level. 
Our results show that for interacting quantum clocks, whether or not non-Markovianity arises due the quantum clock, depends on microscopic clock-system-environment interactions and if the clock coupling operators present in these interactions have quantum fluctuations in the clock states.

Given the results and insights in this work we plan to extend the work in several directions in future research. As discussed in section~\ref{sec:relational-master-equation}, in this work we considered a restricted form of the spectral dressing. This form needs to be generalized to also cover situations like level crossings, certain degeneracies, or qualitative changes of the spectrum such as bound states that appear or disappear. Furthermore, in this work we considered only a single quantum clock and did not address quantum reference frame transformations between different choices of interacting clocks in open quantum systems yet. 
We expect these to be more general for interacting quantum clocks compared to non-interacting ones, thus making this an interesting direction for future work. For recent work on quantum reference frame transformations in open quantum systems with non-interacting clocks, see \cite{Luppi:2026qyv}. 
This question is directly linked to subsystem relativity~\cite{AliAhmad:2021adn,Castro-Ruiz:2021vnq}, since the split into system and environment is frame dependent already for non-interacting clocks ~\cite{AliAhmad:2021adn}. For interacting clocks, this more profound because the interaction Hamiltonian directly couples clock, system and environment. Changing to the perspective of a different quantum clock therefore changes not only the labeling of relational time, but also the effective split into system and environment. For example, the relational master equation might describe pure dephasing from the perspective of one clock, while it might also involve dissipation from the perspective of another, depending on how the clocks couple. 
 
Finally, a longer-term aim is to extend the methods developed in this work to relativistic field theory models while describing the clock at the microscopic level. For recent work in the perspective-neutral field theory models for non-interacting and ideal clocks, see \cite{Freidel:2026gcp} and \cite{Chen:2026kui} respectively. 
This extension is non-trivial due to potential complications arising from the Hilbert space in field theories being non-factorizable into simple tensor products in general. Thus, the spectral dressing requires potentially new methods for field theoretic contexts. This might also provide a way to implement interacting clocks in the field theory models from the more algebraic perspective of crossed products.

\section{Acknowledgements}
KG and AS thank Paul Hartung, Roman Kemper, Max Fahn and Erik Wachter for useful discussions. In addition KG and AS thank Viktoria Kabel and Paolo Luppi for discussion on relational open quantum systems. 
AS thanks the German Academic Scholarship Foundation for financial support. AS further thanks the Friedrich-Naumann Stiftung for funding during the initial phase of this project. AS thanks Perimeter Institute for Theoretical Physics for its hospitality during visits when part of this work was carried out. KG thanks Perimeter Institute for Theoretical Physics for hospitality and support.  KG was supported by a grant from the Simons Foundation (1034867, Dittrich).
The authors acknowledge use of OpenAI's ChatGPT and Anthropic's Claude to assist with transcribing handwritten derivations, literature reviews, algebraic consistency checks, and typesetting the manuscript, including error checking and scripting as well as creating the figures used in this paper. The authors directed the use of these tools, reviewed and verified their outputs, and take full responsibility for the content of this paper.
\appendix
\section{Properties of the POVM $E(\tau)$}
\label{app:povm-properties}
For non-interacting clocks, the clock POVM $E_{\C}(\tau)$ reviewed in subsection~\ref{sec:qclocks-rev} is positive, complete and covariant under $H_{\C}$. As the constraint in \eqref{eq:constraint} does not couple the clock to $\SE$, covariance under $H_{\C}$ is equivalent to covariance under the full constraint. For interacting clocks this is no longer the case, which is why we introduced the dressed POVM $E(\tau)$ in \eqref{eq:rme-povm}. In order for $E(\tau)$ to describe a clock readout in the same way as $E_{\C}(\tau)$ does for non-interacting clocks, it needs to satisfy the same three properties, now with respect to $\CH$. In this appendix, we show that the POVM $E(\tau)$ introduced in \eqref{eq:rme-povm} is positive, normalized, and complete, and that it satisfies the covariance condition~\eqref{eq:rme-covariance}.

Throughout, we denote $\ket{V_A^{(0)}} = \ket{\varepsilon_m^{(0)}}_{\C}\otimes\ket{e_a^{(0)}}_{\SE}$, with label $A\equiv(m,a)$, the eigenstates of $\CO$, and by $\ket{V_A}=\Omega\ket{V_A^{(0)}}$ the eigenstates of $\CH$, i.e., $\CH\ket{V_A} = V_A\ket{V_A}$. Further, $\Pi_A$ denotes the projector onto the eigenspace of $\CH$ with eigenvalue $V_A$, so that $\Pi_A=\Pi_B$ whenever $V_A=V_B$. The clock POVM $E_{\C}(\tau)=\ket{\tau}\bra{\tau}_{\C}$ is the one introduced in subsection~\ref{sec:qclocks-rev} with $\braket{\varepsilon_m^{(0)}|0}_{\C}=1$, and $\frac{1}{\mu}\int \mathrm{d}\tau$ denotes the time average with the same normalization as used there.

{\bfseries Positivity:} Positivity of $E(\tau)$ ensures that the probability $p(\tau)$ of the clock reading the time $\tau$ is non-negative. It follows directly from the construction of $E(\tau)$ in \eqref{eq:rme-povm}. As both $\Omega$ and $e^{-i\tau\CH}$ are unitary, $E(\tau)$ is unitarily equivalent to $E_{\C}(0)\otimes\Id_{\SE}$, which is positive. Explicitly, for any $\ket{\psi}\in\mathcal{H}^\mathrm{kin}$ we define $\ket{\Psi}\coloneqq\Omega^\dagger e^{i\tau\CH}\ket{\psi}$ and obtain
\begin{equation}
  \bra{\psi}E(\tau)\ket{\psi}
  = \bra{\Psi}E_{\C}(0)\otimes\Id_{\SE}\ket{\Psi}
  = \big\|\big(\bra{0}_{\C}\otimes\Id_{\SE}\big)\ket{\Psi}\big\|^2 \geq 0 .
\end{equation}

{\bfseries Normalization and completeness:} Completeness means that the clock readouts form a complete set of outcomes, i.e., averaged over all times the POVM elements add up to the identity, $\frac{1}{\mu}\int \mathrm{d}\tau\:E(\tau)=\Id$. To show this, we use that $E(\tau)$ is constructed via \eqref{eq:rme-povm} as $E(\tau)=e^{-i\tau\CH}\:E(0)\:e^{i\tau\CH}$ with $E(0)\coloneqq\Omega\big(E_{\C}(0)\otimes\Id_{\SE}\big)\Omega^\dagger$. 
Let $\Pi_\lambda$ denote the projector onto the eigenspace of $\CH$ with eigenvalue $\lambda$, such that $\sumint_\lambda\Pi_\lambda=\Id$ and $e^{-i\tau\CH}\Pi_\lambda=e^{-i\tau\lambda}\Pi_\lambda$. Inserting the identity on both sides of $E(0)$, we can write
\begin{equation*}
  E(\tau)
  = \sumint_{\lambda,\lambda'} e^{-i\tau\CH}\:\Pi_\lambda E(0) \Pi_{\lambda'}\:e^{i\tau\CH}
  = \sumint_{\lambda,\lambda'} e^{-i\tau(\lambda-\lambda')}\:\Pi_\lambda E(0) \Pi_{\lambda'} .
\end{equation*}
The only $\tau$-dependence is now in the phases $e^{-i\tau(\lambda-\lambda')}$, and the time average of these gives $\frac{1}{\mu}\int \mathrm{d}\tau\:e^{-i\tau(\lambda-\lambda')}=\delta_{\lambda\lambda'}$. We thus obtain
\begin{align}
  \frac{1}{\mu}\int \mathrm{d}\tau\:E(\tau)
  &= \sumint_\lambda \Pi_\lambda\:\Omega\big(E_{\C}(0)\otimes\Id_{\SE}\big)\Omega^\dagger\:\Pi_\lambda
  \nonumber\\
  &= \sumint_\lambda \sumint_{\substack{A,B\:\mathrm{s.t.}\\ V_A=V_B=\lambda}} \ket{V_A}\bra{V_A^{(0)}}E_{\C}(0)\otimes\Id_{\SE}\ket{V_B^{(0)}}\bra{V_B}\nonumber\\
  &= \sumint_A \ket{V_A}\bra{V_A}\nonumber\\
  &= \Id .
  \label{eq:app-povm-completeness}
\end{align}
In the second equality, we inserted $\Omega=\sumint_A\ket{V_A}\bra{V_A^{(0)}}$. Since $\Pi_\lambda$ projects onto the eigenspace with eigenvalue $\lambda$, we have $\Pi_\lambda\ket{V_A}=\ket{V_A}$ if $V_A=\lambda$ and $\Pi_\lambda\ket{V_A}=0$ otherwise. Hence, $\Pi_\lambda\Omega=\sumint_{A\:\mathrm{s.t.}\:V_A=\lambda}\ket{V_A}\bra{V_A^{(0)}}$, and analogously $\Omega^\dagger\Pi_\lambda=\sumint_{B\:\mathrm{s.t.}\:V_B=\lambda}\ket{V_B^{(0)}}\bra{V_B}$, i.e., only eigenstates with energy $\lambda$ contribute.
For the third equality, note that $E_{\C}(0)$ acts only on the clock, and that the clock state $\ket{0}_{\C}$ has unit overlap with every clock energy eigenstate, $\braket{\varepsilon_m^{(0)}|0}_{\C}=1$. Hence,
\begin{equation*}
  \bra{V_A^{(0)}}E_{\C}(0)\otimes\Id_{\SE}\ket{V_B^{(0)}} = \braket{\varepsilon_m^{(0)}|0}_{\C}\braket{0|\varepsilon_n^{(0)}}_{\C}\:\delta_{ab} = \delta_{ab},
\end{equation*}
i.e., this matrix element only checks whether the $\SE$ labels $a$ and $b$ agree and does not distinguish between the clock labels $m$ and $n$. However, in the second line, $A$ and $B$ always belong to the same eigenspace of $\CH$, namely that corresponding to eigenvalue $\lambda$. As a result, the only pairs that contribute to the double sum/integral in the second line are $A=(m,a)$ and $B=(n,a)$, which share the same $\SE$ label $a$ and, in addition, the same energy, $V_{(m,a)}=V_{(n,a)}=\lambda$. We assume that eigenstates with the same $\SE$ label but different clock labels have different energies, $V_{(m,a)}\neq V_{(n,a)}$ for $m\neq n$. Under this assumption, the two conditions can only be met for $m=n$, i.e., for $A=B$. The double sum/integral therefore reduces to a single sum/integral. Finally performing the remaining sum/integral over $\lambda$ and using completeness of $\Pi_\lambda$ projectors, we get the fourth equality. 
Since the blocks $\Pi_\lambda E(\tau)\Pi_\lambda$ act on mutually orthogonal subspaces, the completeness relation is therefore equivalent to the statement that $\Pi_A\:E(\tau)\:\Pi_A = \Pi_A$ for all $A$ and $\tau$.

{\bfseries Covariance:} Covariance ensures that shifting the clock reading by $\xi$ is the same as evolving along the flow of the full constraint $\CH$ by $\xi$. This is the property that, for non-interacting clocks, was guaranteed by covariance of $E_{\C}(\tau)$ under $H_{\C}$, and which for interacting clocks has to hold with respect to $\CH$ instead. Since $E(\tau)$ in \eqref{eq:rme-povm} is defined by evolving $E(0)$ with $\CH$, covariance holds by construction:
\begin{equation}
  E(\tau+\xi)
  = e^{-i\xi\CH}\,e^{-i\tau\CH}\,\Omega\big(E_{\C}(0)\otimes\Id_{\SE}\big)\Omega^\dagger\,e^{i\tau\CH}\,e^{i\xi\CH}
  = e^{-i\xi\CH}\,E(\tau)\,e^{i\xi\CH},
\end{equation}
which is \eqref{eq:rme-covariance}. Differentiating with respect to $\xi$ at $\xi=0$ yields the equivalent infinitesimal form $\partial_\tau E(\tau)=-i[\CH,E(\tau)]$.

\section{Exact master equation for the qubit clock and system}
\label{app:exact-qubit-result}
For the qubit clock model of subsection~\ref{sec:ex-qubit}, the full spectrum of $\CH$ is known, such that the relational state and the kernel \eqref{eq:Def-kernel-tau} can be computed exactly. For $\varphi=0$ and as long as $\cos(2\beta\tau/\hbar)\neq 0$, the exact relational master equation can be derived to be
\begin{align*}
  \frac{\mathrm{d}}{\mathrm{d}\tau}\rho_\S(\tau)
  &= -\frac{i}{\hbar}\Big[ H_\S + \frac{\alpha}{\cos\theta}\cos \big(\tfrac{2\beta\tau}{\hbar}\big)\: \widetilde{X}_\S\big(\tfrac{\zeta}{\hbar\omega}\tau\big),\: \rho_\S(\tau) \Big]\\
  &\hspace{1.5cm} 
  + \frac{\zeta-\hbar\omega}{2\hbar}\: \Big[\widetilde{Y}_\S\big(\tfrac{\zeta}{\hbar\omega}\tau\big), \big[\widetilde{X}_\S\big(\tfrac{\zeta}{\hbar\omega}\tau\big), \rho_\S(\tau)\big]\Big]
  \\
  &\hspace{1.5cm} - \frac{\beta}{2\hbar}\tan \big(\tfrac{2\beta\tau}{\hbar}\big)\: \big[Z_\S, [Z_\S, \rho_\S(\tau)]\big]\\
  &\hspace{1.5cm} 
  - \frac{\alpha}{2\hbar\cos\theta}\sin \big(\tfrac{2\beta\tau}{\hbar}\big)\tan \big(\tfrac{2\beta\tau}{\hbar}\big)\: \Big[\widetilde{Y}_\S\big(\tfrac{\zeta}{\hbar\omega}\tau\big), [Z_\S, \rho_\S(\tau)]\Big]
  \\
  &\hspace{1.5cm} + \frac{i\alpha}{2\hbar\cos\theta}\sin \big(\tfrac{2\beta\tau}{\hbar}\big)\: \Big[\widetilde{Y}_\S\big(\tfrac{\zeta}{\hbar\omega}\tau\big), \{Z_\S, \rho_\S(\tau)\}\Big]\\
  &\hspace{1.5cm} 
  + \frac{i\beta\sin\theta}{2\hbar\cos \big(\tfrac{2\beta\tau}{\hbar}\big)}\: \Big[Z_\S, \big\{\widetilde{X}_\S\big(\tfrac{\zeta}{\hbar\omega}\tau\big), \rho_\S(\tau)\big\}\Big]\\
  &\hspace{1.5cm} - \frac{i(\zeta-\hbar\omega)}{2\hbar}\tan \big(\tfrac{2\beta\tau}{\hbar}\big)\: \Big[\widetilde{X}_\S\big(\tfrac{\zeta}{\hbar\omega}\tau\big), \big\{\widetilde{Y}_\S\big(\tfrac{\zeta}{\hbar\omega}\tau\big), \rho_\S(\tau)\big\}\Big],
\end{align*}
where $\theta=\frac{1}{2}\arctan \big(\alpha/(\hbar\omega)\big)$ and $\zeta=\sqrt{(\hbar\omega)^2+\alpha^2}$. The operators $\widetilde{X}_\S$ and $\widetilde{Y}_\S$ are the same as in subsection~\ref{sec:ex-qubit}, but evaluated at the rescaled time $\frac{\zeta}{\hbar\omega}\tau$, i.e., they rotate with the frequency $\zeta/\hbar$ of the full constraint instead of $\omega$. For a general phase choice $\varphi$, the arguments of all interaction operators are shifted by $\frac{\varphi}{2\omega}$, while the coefficients remain unchanged.

If we expand this exact master equation to second order jointly in $\alpha$ and $\beta$, we get
\begin{alignat*}{2}
  \frac{\alpha}{\cos\theta}\cos \big(\tfrac{2\beta\tau}{\hbar}\big)
  &= \alpha\Big(1+\frac{\alpha^2}{8\hbar^2\omega^2}+\dots\Big)\Big(1-\frac{2\beta^2\tau^2}{\hbar^2}+\dots\Big)
  &&\approx \alpha,
  \\
  \frac{\zeta-\hbar\omega}{2\hbar}
  &= \frac{1}{2\hbar}\Big(\frac{\alpha^2}{2\hbar\omega}-\frac{\alpha^4}{8\hbar^3\omega^3}+\dots\Big)
  &&\approx \frac{\alpha^2}{4\omega\hbar^2},
  \\
  \frac{\beta}{2\hbar}\tan \big(\tfrac{2\beta\tau}{\hbar}\big)
  &= \frac{\beta}{2\hbar}\Big(\frac{2\beta\tau}{\hbar}+\frac{8\beta^3\tau^3}{3\hbar^3}+\dots\Big)
  &&\approx \frac{\beta^2\tau}{\hbar^2},
  \\
  \frac{\alpha}{2\hbar\cos\theta}\sin \big(\tfrac{2\beta\tau}{\hbar}\big)
  &= \frac{\alpha}{2\hbar}\Big(1+\frac{\alpha^2}{8\hbar^2\omega^2}+\dots\Big)\Big(\frac{2\beta\tau}{\hbar}-\frac{4\beta^3\tau^3}{3\hbar^3}+\dots\Big)
  &&\approx \frac{\alpha\beta\tau}{\hbar^2},
  \\
  \frac{\beta\sin\theta}{2\hbar\cos \big(\tfrac{2\beta\tau}{\hbar}\big)}
  &= \frac{\beta}{2\hbar}\Big(\frac{\alpha}{2\hbar\omega}+\dots\Big)\Big(1+\frac{2\beta^2\tau^2}{\hbar^2}+\dots\Big)
  &&\approx \frac{\alpha\beta}{4\omega\hbar^2},
  \\
  \frac{\alpha}{2\hbar\cos\theta}\sin \big(\tfrac{2\beta\tau}{\hbar}\big)\tan \big(\tfrac{2\beta\tau}{\hbar}\big)
  &= \frac{\alpha}{2\hbar}\Big(\frac{2\beta\tau}{\hbar}\Big)^2+\dots
  &&\approx 0,
  \\
  \frac{\zeta-\hbar\omega}{2\hbar}\tan \big(\tfrac{2\beta\tau}{\hbar}\big)
  &= \frac{\alpha^2}{4\omega\hbar^2}\:\frac{2\beta\tau}{\hbar}+\dots
  &&\approx 0.
\end{alignat*}

For $\alpha=0$, we get $\theta=0$ and $\zeta=\hbar\omega$. The exact master equation then reduces to
\begin{equation*}
  \frac{\mathrm{d}}{\mathrm{d}\tau}\rho_\S(\tau)=-\frac{i}{\hbar}\big[H_\S,\rho_\S(\tau)\big]-\frac{\beta}{2\hbar}\tan \big(\tfrac{2\beta\tau}{\hbar}\big)\:\big[Z_\S,[Z_\S,\rho_\S(\tau)]\big].
\end{equation*}
which describes pure dephasing in the eigenbasis of $Z_\S$. The populations $\bra{\uparrow}\rho_\S(\tau)\ket{\uparrow}$ and $\bra{\downarrow}\rho_\S(\tau)\ket{\downarrow}$ remain constant, while the coherences evolve as
\begin{equation*}
\bra{\uparrow}\rho_\S(\tau)\ket{\downarrow}=e^{-2i\omega\tau}\cos \big(\tfrac{2\beta\tau}{\hbar}\big)\:\bra{\uparrow}\rho_\S(0)\ket{\downarrow},
\end{equation*}
Hence, on top of the free unitary evolution, the coherences are reduced by the factor $\cos \big(\tfrac{2\beta\tau}{\hbar}\big)$. The interaction $\beta\:Z_\C\otimes Z_\S$ shifts the energy splitting of $\S$ by $+2\beta$ or $-2\beta$ depending on whether $Z_\C=+1$ or $Z_\C=-1$, such that the coherence acquires the phase $e^{-2i\beta\tau/\hbar}$ or $e^{+2i\beta\tau/\hbar}$. Since every clock state is an equal superposition of both eigenstates of $Z_\C$, the relational state contains the average of the two phases, $\frac{1}{2}\big(e^{-2i\beta\tau/\hbar}+e^{2i\beta\tau/\hbar}\big)=\cos \big(\tfrac{2\beta\tau}{\hbar}\big)$. This dephasing is a genuine quantum clock effect, since it originates from $Z_\C$ not taking a sharp value in the clock states.

In comparison, the second order master equation of subsection~\ref{sec:ex-qubit} for $\alpha=0$ contains the term $-\frac{\beta^2\tau}{\hbar^2}[Z_\S,[Z_\S,\rho_\S]]$ instead, which yields
\begin{equation*}
\bra{\uparrow}\rho_\S(\tau)\ket{\downarrow}=e^{-2i\omega\tau}\:e^{-2\beta^2\tau^2/\hbar^2}\:\bra{\uparrow}\rho_\S(0)\ket{\downarrow},
\end{equation*}
i.e., a Gaussian decay of the coherences. Both results agree for $\beta\tau/\hbar\ll1$, since $\cos \big(\tfrac{2\beta\tau}{\hbar}\big)=1-\frac{2\beta^2\tau^2}{\hbar^2}+\dots$ and $e^{-2\beta^2\tau^2/\hbar^2}=1-\frac{2\beta^2\tau^2}{\hbar^2}+\dots$, but they differ at later times. While the Gaussian decays monotonically, the exact factor $\cos\big(\tfrac{2\beta\tau}{\hbar}\big)$ reaches zero at $\tau=\pi\hbar/(4\beta)$ and afterwards becomes non-zero again.

At $\tau=\pi\hbar/(4\beta)$, the coherences vanish for every initial state, such that initial states that differ only in their coherences lead to the same relational state. Hence, the map $\Lambda_\tau$ cannot be inverted at this time, and correspondingly the coefficient $\tan\big(\tfrac{2\beta\tau}{\hbar}\big)$ of the exact master equation diverges. This is the exact counterpart of the invertibility condition on $\Lambda_\tau$ discussed in section~\ref{sec:relational-master-equation}.

\section{Exact TCL equation for the qubit clock and system in the Kossakowski form}
\label{app:exact-qubit-kossakowski}
The exact TCL equation from appendix~\ref{app:exact-qubit-result} for the qubit example can be rewritten and recast in the so-called \textit{Kossakowski form}~\cite{Gorini:1975nb,Hall:2014oiy} since its generator is time-local and preserves Hermiticity and trace. This form reads
\begin{equation}
  \frac{\mathrm{d}}{\mathrm{d}\tau}\rho_\S(\tau)
  = -\frac{i}{\hbar}\big[H_{\rm eff}(\tau),\rho_\S(\tau)\big]
  + \sum_{i,j=1}^{3}\Gamma_{ij}(\tau)\Big(A_i\,\rho_\S(\tau)\,A_j-\frac{1}{2}\big\{A_jA_i,\rho_\S(\tau)\big\}\Big),
  \label{eq:kossakowski}
\end{equation}
where $H_{\rm eff}(\tau)$ is an effective Hamiltonian, $\Gamma(\tau)$ is the Hermitian $3\times3$ \textit{Kossakowski matrix}, and $A_1$, $A_2$, $A_3$ are traceless self-adjoint mutually orthogonal operators on $\mathcal{H}_\S$. The eigenvalues of the Kossakowski matrix $\Gamma(\tau)$ give us the decoherence rates of the independent decoherence quantum channels of which there are as many as the rank of the $\Gamma(\tau)$ matrix. The dynamics is said to be CP-divisible if and only if all eigenvalues of $\Gamma(\tau)$ are non-negative for all $\tau$~\cite{Hall:2014oiy, Plenio:2014nyj}. Note, that this is not the Lindblad form mentioned in subsection~\ref{sec:OQS}, which requires the Kossakowski matrix $\Gamma(\tau)$ to be constant in time and positive semidefinite, i.e., all its eigenvalues to be non-negative.

For the exact TCL from appendix~\ref{app:exact-qubit-result} for the qubit example, the Kossakowski operators can be obtained to be
\begin{align*}
  A_1 &= \widetilde{X}_\S\big(\tfrac{\zeta}{\hbar\omega}\tau\big),\\
  A_2 &= \widetilde{Y}_\S\big(\tfrac{\zeta}{\hbar\omega}\tau\big),\\
  A_3 &= Z_\S,
\end{align*}
with $\widetilde{X}_\S$ and $\widetilde{Y}_\S$ as in appendix~\ref{app:exact-qubit-result}.
Expanding the nested commutators and anticommutators in the exact TCL, the effective Hamiltonian can be obtained to be
\begin{equation*}
  H_{\rm eff}(\tau) = H_\S+\frac{\zeta-\hbar\omega}{2}\,Z_\S
  +\frac{\alpha}{2\cos\theta}\bigg(\cos\big(\tfrac{2\beta\tau}{\hbar}\big)+\frac{1}{\cos\big(\tfrac{2\beta\tau}{\hbar}\big)}\bigg)\,\widetilde{X}_\S\big(\tfrac{\zeta}{\hbar\omega}\tau\big)
\end{equation*}
 where $\theta$ and $\zeta$ have the same definitions as earlier. This contains the first-order shift to the Hamiltonian from TCL$_2$ equation for the qubit example as well as a second order Lamb shift-type correction that was hidden inside the second order dissipator. The Kossakowski matrix can be obtained to be
\begin{equation*}
  \Gamma(\tau)=
  \begin{pmatrix}
    \Gamma_{11}(\tau) & \Gamma_{12}(\tau) & \Gamma_{13}(\tau)\\
    \Gamma_{12}^*(\tau) & \Gamma_{22}(\tau) & \Gamma_{23}(\tau)\\
    \Gamma_{13}^*(\tau) & \Gamma_{23}^*(\tau) & \Gamma_{33}(\tau)
  \end{pmatrix},
\end{equation*}
with
\begin{align*}
  \Gamma_{12}(\tau) &= -\frac{\zeta-\hbar\omega}{2\hbar}\Big(1+i\tan\big(\tfrac{2\beta\tau}{\hbar}\big)\Big),\\
  \Gamma_{13}(\tau) &= -\frac{i\beta\sin\theta}{2\hbar\cos\big(\tfrac{2\beta\tau}{\hbar}\big)},\\
  \Gamma_{23}(\tau) &= \frac{\alpha}{2\hbar\cos\theta}\sin\big(\tfrac{2\beta\tau}{\hbar}\big)\Big(\tan\big(\tfrac{2\beta\tau}{\hbar}\big)+i\Big),\\
  \Gamma_{33}(\tau) &= \frac{\beta}{\hbar}\tan\big(\tfrac{2\beta\tau}{\hbar}\big),\\
  \Gamma_{11}(\tau)&=\Gamma_{22}(\tau)=0.
\end{align*}
From this, upon diagonalization, one sees that one of the eigenvalues of the $\Gamma(\tau)$ matrix vanishes, i.e., $\gamma_0=0$ while the two remaining eigenvalues $\gamma_\pm$ are
\begin{equation*}
  \gamma_\pm(\tau) = \frac{\Gamma_{33}(\tau)}{2}\pm\sqrt{\frac{\Gamma_{33}(\tau)^2}{4}+|\Gamma_{12}(\tau)|^2+|\Gamma_{13}(\tau)|^2+|\Gamma_{23}(\tau)|^2},
\end{equation*}
and they satisfy $\gamma_+\gamma_- = -\big(|\Gamma_{12}|^2+|\Gamma_{13}|^2+|\Gamma_{23}|^2\big)$. Therefore, whenever any off-diagonal entry in $\Gamma$ is non-vanishing, the expression under the square root in $\gamma_\pm(\tau)$ is larger than the first term. Thus one gets $\gamma_+(\tau)>0$ and $\gamma_-(\tau)<0$.
The negative decoherence rate $\gamma_-(\tau)$ means that $\Gamma(\tau)$ is not positive semidefinite, such that the exact relational dynamics of $\S$ is not CP-divisible for the qubit example. 

\renewcommand{\bibsection}{\section*{References}}
\bibliographystyle{unsrtnat}
\bibliography{references}

\end{document}